\documentclass{aa}
\usepackage{multirow}
\usepackage{epsfig}
\usepackage{subfigure}
\usepackage{txfonts}
\usepackage{tikz}
\usepackage{xcolor}
\usepackage{wasysym}
\usepackage{amssymb}
\usepackage{multicol}
\usepackage{lscape}
\usepackage{hyperref}
\usepackage{graphicx}

\makeatletter
\let\jnl@style=\rm
\def\ref@jnl#1{{\jnl@style#1}}

\def\aj{\ref@jnl{AJ}}                   
\def\araa{\ref@jnl{ARA\&A}}             
\def\apj{\ref@jnl{ApJ}}                 
\def\apjl{\ref@jnl{ApJ}}                
\def\apjs{\ref@jnl{ApJS}}               
\def\ao{\ref@jnl{Appl.~Opt.}}           
\def\apss{\ref@jnl{Ap\&SS}}             
\def\aap{\ref@jnl{A\&A}}                
\def\aapr{\ref@jnl{A\&A~Rev.}}          
\def\aaps{\ref@jnl{A\&AS}}              
\def\azh{\ref@jnl{AZh}}                 
\def\baas{\ref@jnl{BAAS}}               
\def\jrasc{\ref@jnl{JRASC}}             
\def\memras{\ref@jnl{MmRAS}}            
\def\mnras{\ref@jnl{MNRAS}}             
\def\pra{\ref@jnl{Phys.~Rev.~A}}        
\def\prb{\ref@jnl{Phys.~Rev.~B}}        
\def\prc{\ref@jnl{Phys.~Rev.~C}}        
\def\prd{\ref@jnl{Phys.~Rev.~D}}        
\def\pre{\ref@jnl{Phys.~Rev.~E}}        
\def\prl{\ref@jnl{Phys.~Rev.~Lett.}}    
\def\pasp{\ref@jnl{PASP}}               
\def\pasj{\ref@jnl{PASJ}}               
\def\rmxaa{\ref@jnl{RMXAA}}             
\def\qjras{\ref@jnl{QJRAS}}             
\def\skytel{\ref@jnl{S\&T}}             
\def\solphys{\ref@jnl{Sol.~Phys.}}      
\def\sovast{\ref@jnl{Soviet~Ast.}}      
\def\ssr{\ref@jnl{Space~Sci.~Rev.}}     
\def\zap{\ref@jnl{ZAp}}                 
\def\nat{\ref@jnl{Nature}}              
\def\iaucirc{\ref@jnl{IAU~Circ.}}       
\def\aplett{\ref@jnl{Astrophys.~Lett.}} 
\def\apspr{\ref@jnl{Astrophys.~Space~Phys.~Res.}}
\def\bain{\ref@jnl{Bull.~Astron.~Inst.~Netherlands}}
\def\fcp{\ref@jnl{Fund.~Cosmic~Phys.}}  
\def\gca{\ref@jnl{Geochim.~Cosmochim.~Acta}}   
\def\grl{\ref@jnl{Geophys.~Res.~Lett.}} 
\def\jcp{\ref@jnl{J.~Chem.~Phys.}}      
\def\jgr{\ref@jnl{J.~Geophys.~Res.}}    
\def\jqsrt{\ref@jnl{J.~Quant.~Spec.~Radiat.~Transf.}}
\def\memsai{\ref@jnl{Mem.~Soc.~Astron.~Italiana}}
\def\nphysa{\ref@jnl{Nucl.~Phys.~A}}   
\def\physrep{\ref@jnl{Phys.~Rep.}}   
\def\physscr{\ref@jnl{Phys.~Scr}}   
\def\planss{\ref@jnl{Planet.~Space~Sci.}}   
\def\procspie{\ref@jnl{Proc.~SPIE}}   

\makeatother

\newcommand {\apgt} {\ {\raise-.5ex\hbox{$\buildrel>\over\sim$}}\ }
\newcommand {\aplt} {\ {\raise-.5ex\hbox{$\buildrel<\over\sim$}}\ } 
\begin{document}
\title{Breaking the rules at {\it z}$\simeq$0.45: the rebel case of RBS 1055}
\author{A. Marinucci
    \inst{1},        
          G. Vietri\inst{2,3},
          E. Piconcelli\inst{2},
          S. Bianchi\inst{4},
          M. Guainazzi\inst{5},\\
          G. Lanzuisi\inst{6},
          D. Stern\inst{7}
          \and
          C. Vignali\inst{8,6}
          }
   \institute{ASI - Italian Space Agency, Via del Politecnico snc, 00133, Rome, Italy
              \email{andrea.marinucci@asi.it}
         \and
              INAF - Osservatorio Astronomico di Roma, via Frascati 33, 00040, Monte Porzio Catone, Roma, Italy
         \and
         INAF - Istituto di Astrofisica Spaziale e Fisica cosmica Milano, via A. Corti 12, 20133, Milano, Italy
          \and
             Dipartimento di Matematica e Fisica, Universit\`a degli Studi Roma Tre, via della Vasca Navale 84, 00146 Roma, Italy
          \and
             ESA - European Space Research and Technology Centre (ESTEC), Keplerlaan 1, 2201 AZ, Noordwijk, The Netherlands
          \and
             INAF - Osservatorio di Astrofisica e Scienza dello Spazio, via P. Gobetti 93/3, 40129, Bologna, Italy
           \and
             Jet  Propulsion  Laboratory,  California  Institute  of  Technology, 4800 Oak Grove Drive, MS 169-224, Pasadena, CA 91109, USA
            \and
             Dipartimento di Fisica e Astronomia “Augusto Righi”, Universit\`{a} degli Studi di Bologna, via P. Gobetti 93/2, 40129 Bologna, Italy
             }
             

 
  \abstract
  {Very luminous quasars are unique sources for studying the circumnuclear environment around supermassive black holes. 
Several components contribute to the overall X-ray spectral shape of Active Galactic Nuclei (AGN). The hot (kT$_e$=50-100 keV) and warm (kT$_e$=0.1-1 keV) coronae are responsible for the hard and soft power law continua, while the circumnuclear toroidal reflector accounts for the Fe K$\alpha$ emission line and the associated Compton hump. However, all these spectral features are simultaneously observed only in a handful of sources above $z\simeq$0.1.}
   {An ideal astrophysical laboratory for this investigation is the quasar RBS 1055, at z$\simeq$0.45. With a luminosity L${}_{\rm 2-10\ keV}=2\times10^{45}$ erg s$^{-1}$, it is the brightest radio-quiet quasar from the ROSAT Bright Survey. Despite the known anti-correlation between the Equivalent Width (EW) of the narrow neutral Fe K$\alpha$ line and L${}_{\rm 2-10\ keV}$, an intense Fe K$\alpha$ was previously detected for this source. }
  {We report on a long (250 ks) {\it NuSTAR} observation performed in March 2021 and archival XMM-{\it Newton} pointings (185 ks) taken in July 2014. An optical spectrum of the source taken with the Double Spectrograph at the Palomar Observatory, quasi-simultaneous with the {\it NuSTAR} observations, is also analyzed.}
   {We find that the two-coronae model, in which a warm and hot corona coexist, well reproduces the broad band spectrum of RBS 1055, with temperatures kT$_e=0.12^{+0.08}_{-0.03}$ keV, kT$_e=30^{+40}_{-10}$ keV and Thomson optical depths $\tau$=30$_{-10}^{+15}$ and  $\tau$=3.0$_{-1.4}^{+1.0}$ for the former and the latter component, respectively. We confirm the presence of an intense Fe K$\alpha$ emission line (EW=55$\pm$6 eV) and, when a toroidal model is considered for reproducing the Compton reflection, a Compton-thin solution with N$_{\rm H}=(3.2^{+0.9}_{-0.8})\times10^{23}$ cm$^{-2}$ for the circumnuclear reflector is found. A detailed analysis of the optical spectrum reveals a likely peculiar configuration of our line of sight with respect to the nucleus, and the presence of a broad [O III] component, tracing outflows in the NLR, with a velocity shift $v=$1500$\pm100$ km s$^{-1}$, leading to a mass outflow rate $\dot{M}_{\rm out}=25.4\pm1.5$ M$_{\odot}$ yr$^{-1}$ and outflow kinetic power normalized by the bolometric luminosity $\dot{E}_{\rm kin}$/L$_{\rm Bol}$ $\sim$0.33\%. We estimate the BH mass to be in the range 2.8$\times$10$^{8}$-1.2$\times$10$^{9}$ M$_{\odot}$, according to different BLR emission lines, with an average value of <$M\rm_{BH}$>=6.5$\times$10$^{8}$ M$_{\odot}$.}
   {With an Fe K$\alpha$ which is 3$\sigma$ above the value predicted from the EW-L$_{2-10\ \rm keV}$ relation and an extreme source brightness at 2 keV (a factor 10-15 higher than the one expected from the optical/UV), RBS 1055 confirms to be an outlier in the X-rays, compared to other objects in the same luminosity and redshift range.}

   \keywords{Galaxies: active --
	      Galaxies: accretion --
	      Individual: RBS 1055
               }
\titlerunning{The rebel case of RBS 1055 at $z\simeq$0.45}
\authorrunning{A. Marinucci, et al.}
   \maketitle       

\section{Introduction}
\indent The commonly accepted paradigm for luminous Active Galactic Nuclei (AGN) postulates the existence of a supermassive black hole at their center, surrounded by an accretion disk and a hot cloud of electrons ($kT_e\sim$50--100 keV) which are responsible for the nuclear continuum. UV/optical seed photons coming from the accretion disk illuminate the corona and are scattered, via the inverse Compton mechanism,  up to the X-ray band \citep[the so called two-phase model:][]{hm91,hmg94}. The slope of the nuclear X-ray power-law continuum is a function of the plasma temperature kT$_e$ and optical depth $\tau$, while the cutoff energy (E$_{\rm C}$) is mainly related to the former parameter \citep{sle76, rl79,st80, lz87, bel99, petr00, phm01}.  Measuring the photon index and the cutoff energy of the primary power-law is fundamental for determining the properties of the hot corona. \\
\indent
{\it NuSTAR} \citep{nustar}, with its broad spectral coverage, has led to a number of studies that have shown that the primary emission of Seyfert galaxies can be well approximated as a cutoff power law with a photon index $\Gamma=1.7$--2.0 and a high energy rollover at E$_{\rm C}$=50--300 keV \citep{flk15, flb17, tbm18, mbm19, bhm20}.  At higher luminosities, the high energy cutoffs measured in the four most distant quasars with L$_{2-10\ {\rm keV}}>10^{45}$ erg s$^{-1}$ are $106^{+102}_{-37}$ keV and $66^{+17}_{-12}$ keV in  2MASSJ1614346+470420 and B1422+231 \citep{lgc19}, $99^{+67}_{-35}$ keV in B2202-209 \citep{krs17} and $99^{+91}_{-35}$ keV in APM 08279+5255 \citep{bvl22}. However, the reprocessing of such a continuum radiation of the AGN from the circumnuclear material makes the measurement of the photon index and of the cutoff energy degenerate with other parameters, like the amount of radiation reflected by circumnuclear matter, which produces a Compton hump at $\sim30$ keV \cite{mpp91,gf91}. The Compton reflection component is accompanied by fluorescent lines emitted from metals, among which the most prominent one is the neutral Fe K$\alpha$ line at 6.4 keV. The strength of the reflection component is traditionally measured as the solid angle subtended by a plane-parallel reflector \citep[in units of 2$\pi$; $R$][]{mz95}. Thanks to their high X-ray fluxes, local (z $\ll$0.1) AGN at moderate luminosity (L$_{2-10\ {\rm keV}}\approx10^{42-43}$ erg s$^{-1}$), have been deeply observed by X-ray facilities, and our knowledge of physical and spectral properties of the X-ray emitting/absorbing regions in AGN have been derived from this class of sources. Studies based on {\it Beppo}SAX \citep{per02}, {\it INTEGRAL} \citep{lub16}, {\it Suzaku} \citep{kut16}, Swift/BAT \citep{rtk17} and {\it NuSTAR} data \citep[][and references therein]{bal18} have found that the bulk of the Seyfert galaxy population exhibit $0.5 \leq R \leq 1.5$. \\
\indent On the contrary, the properties of the hard X-ray emission of more luminous AGN have remained poorly investigated because they are rare in the local Universe, and - while intrinsically more luminous - their observed flux is much dimmer than for nearby Seyfert galaxies. X-ray observational programs targeting luminous AGN are time-consuming, and therefore scarce. This is even more true for the reflection which substantially arises and reaches its maximum above 10 keV, due to the well-known instrumental limitations in terms of imaging and spectroscopy affecting past X-ray missions in this high energy range. Measurements of $R$ for luminous AGN are indeed very scarce and poorly constrained, being basically derived from the 0.5-10 keV spectra, and seem to suggest low values, i.e. $R\ll1$ \citep{rt00,page05}, even when {\it NuSTAR} observations are taken into account \citep{zcc18}. The claim for a weak reflection component in luminous AGN is mainly supported by the observed anti-correlation between the Equivalent Width (EW) of the narrow neutral Fe K$\alpha$ line and L$_{2-10\ {\rm keV}}$ \citep[e.g. the IT effect;][]{it93,page04,bianchi07, jiang06, shu10, shu12, ricci14}. The IT effect is interpreted in terms of a decreasing covering factor of the Compton-thick torus (i.e. the main reflector) as a function of increasing L$_{2-10\ {\rm keV}}$. However, the quality of the data for objects with L$_{2-10\ {\rm keV}}>10^{45}$ erg s$^{-1}$ is typically too poor even to put tight constraints on the EW of the Fe K$\alpha$ line \citep{jpg05}. The need for good measurements of $R$ in luminous AGN is thus compelling to accurately confirm the weakness/absence of the reflection component in their high-energy X-ray spectra.\\
\indent A further spectral component has also been invoked in the past to reproduce the soft excess of AGN \citep[i.e. photons in the 0.5-2 keV band in excess of the extrapolation of the hard power-law component:][]{arn85,sgn85}. These models assume a thermal Comptonisation in an optically thick ($\tau$=5-50) and warm (kT$_e$=0.1-1 keV) corona \citep{mbz98, pro04, ddj12, jwd12, ppm13, rmb15} and have been tested on a handful of sources above $z=$0.1 \citep{pud18}. \\
\indent An excellent candidate to study the nuclear and circumnuclear environment in highly luminous AGN is RBS 1055, at z$\simeq$0.45. With a luminosity L${}_{\rm 2-10}=2\times10^{45}$ erg s$^{-1}$, it is the brightest radio-quiet quasar from the ROSAT Bright Survey \citep[RBS:][]{shl00}, as reported in \citet{klm10}, excluding radio-loud objects since their X-ray emission can be heavily contaminated by the relativistic jet-related component. The black hole mass estimate for this source is M$_{\rm BH}=7.5\times10^8$ M$_{\odot}$ \citep{woo08}. RBS 1055 was observed by XMM-{\it Newton} in 2008 for a net observing time of 21 ks, from which an Fe K$\alpha$ EW of $140^{+70}_{-50}$ eV, as well as a strong soft-excess below $\sim$1 keV were detected \citep{klm10}. A comparison with other sources in the literature reveals that this is the most significant EW measurement above $10^{45}$ erg s$^{-1}$ \citep{bianchi07}.  \\
\indent In this work, we analyze novel {\it NuSTAR} observations (250 ks) performed in March 2021 and past archival XMM-{\it Newton} pointings (185 ks) taken in July 2014. The source was also observed with the Double Spectrograph at the Palomar Observatory on March 2021, quasi-simultaneously with {\it NuSTAR}, and we infer a more refined redshift measurement $z=0.452\pm0.001$ (Sect. 4), which we will use in this paper. Data from the Optical Monitor (OM) on board XMM-{\it Newton} allow us to test the two-coronae model in a source at redshift $z\simeq$0.45. We adopt the cosmological parameters $H_0=70$ km s$^{-1}$ Mpc$^{-1}$, $\Omega_\Lambda=0.73$ and $\Omega_m=0.27$ throughout the paper, i.e. the default ones in \textsc{xspec 12.12.1} \citep{xspec}. Errors correspond to the 90\% confidence level for one interesting parameter ($\Delta\chi^2=2.7$), if not stated otherwise. 

\begin{figure*}
\centering
  \epsfig{file=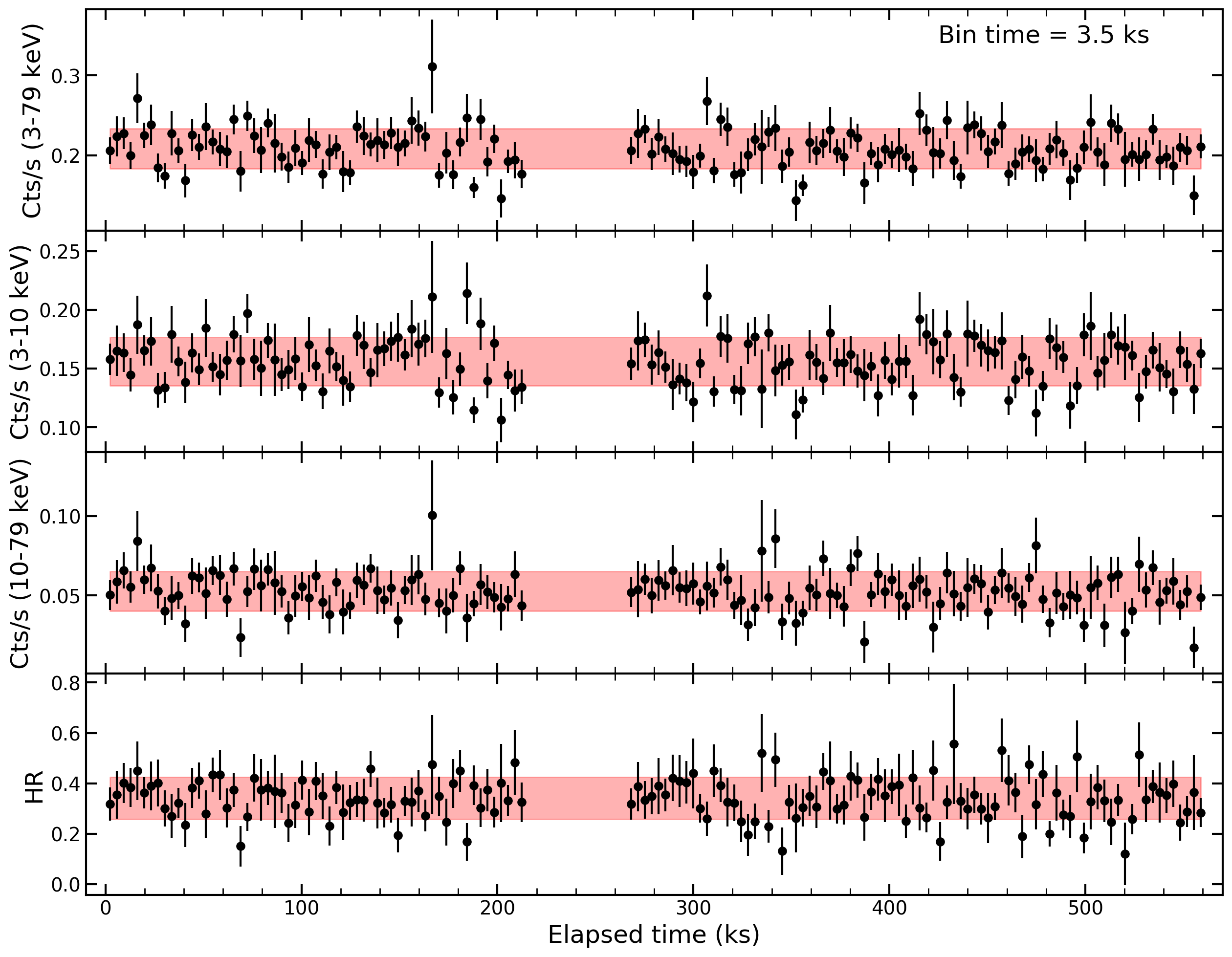, width=0.8\textwidth}
  \caption{{\it NuSTAR} FPMA+B light curves in the 3-79 keV, 3-10 keV and 10-79 keV energy bands are shown in the top panels, with a binning time of 3.5 ks. The hardness ratio is shown in the bottom panel. Red shaded regions indicate mean count rates and hardness ratio $\pm1\sigma$.}
  \label{fig:nulc}
\end{figure*}

\section{Observations and data reduction}
\subsection{XMM-Newton}
RBS 1055 has been targeted by XMM-{\it Newton} \citep{xmm} three times, on 2008 November 6 for a total elapsed time of $24.9$ ks \citep[ObsID 0555020201: reported in][]{klm10}, on 2014 July 13 for a total elapsed time of $138.7$ ks (ObsID 0744450101) and on 2014 July 15 for a total elapsed time of $67.7$ ks (ObsID 0744450401). These observations used the EPIC CCD cameras: the pn \citep{struder01} and the two MOS \citep{turner01}, operated in full/large window and thin filter mode. We analyze only data from the last two observations in this paper, due to the higher Signal-to-Noise Ratio (SNR) throughout the 0.5-10 keV energy band. This data set has not been published yet. The extraction radii and the optimal time cuts for flaring particle background were computed with SAS 19 \citep{gabr04} via an iterative process which leads to a maximization of the SNR, similar to the approach described in \citet{pico04}. The resulting optimal extraction radii are 40 arcsec, net exposure times are 123.6 ks and 60.5 ks for the pn spectra and 137.7 ks and 64.5 ks for the summed MOS1+2 spectra, respectively. Background spectra were extracted from source-free circular regions with a radius of 50 arcsec. Spectra were then binned in order not to over-sample the instrumental resolution by more than a factor of three and to have no less than 30 counts in each background-subtracted spectral channel. Since no significant variability in spectral shape and flux is observed within each observation, we coadded the two data sets. The final net exposure times are therefore 184 ks and 202 ks for the pn and the MOS1+2 spectra. A cross-calibration factor within five percent between the two spectra is found, this is taken into account via the adoption of a multiplicative constant in the model. The two XMM-{\it Newton} pointings also have data from the Optical Monitor, with the UVM2 (2310 \AA), UVW1(2910 \AA) and U (3440 \AA) filters. We reduced the data with the {\sc omichain} tool and constructed {\sc Xspec} readable spectra with {\sc om2pha}. Since no statistically significant difference is found between the two pointings, we used spectra from the longest one, i.e. {ObsID 0744450101}.
\subsection{NuSTAR}
{\it NuSTAR} \citep{nustar} observed RBS 1055 with its two coaligned X-ray telescopes with corresponding Focal Plane Module A (FPMA) and B (FPMB) simultaneously on 2021 March 3 and 6 for a total of $212$ ks and 293 ks of elapsed time, respectively.  The Level 1 data products were processed with the {\it NuSTAR} Data Analysis Software (NuSTARDAS) package (v. 2.0.0). Cleaned event files (level 2 data products) were produced and calibrated using standard filtering criteria with the \textsc{nupipeline} task and the calibration files available in the {\it NuSTAR} calibration database (CALDB 20210202). Extraction radii for the source and background spectra were $30$ arcsec and 60 arcsec and the net exposure times for the two observations were 108.7/107.7 ks and 145.6/144.2 ks for FPMA/B, respectively.  The two pairs of {\it NuSTAR} spectra were binned in order not to over-sample the instrumental resolution by more than a factor of 2.5 and to have a SNR  higher than 3 in each spectral channel. A cross-calibration factor within three percent between the two detectors is found. Summed FPMA/B light curves in different energy bands are shown in Fig. \ref{fig:nulc}. Red shaded regions indicate mean count rates and hardness ratio $\pm1$ standard deviation. Since no spectral or flux variations are found, the FPMA/B spectra from the two observations are summed, for a net observing time of 254.3 ks and 251.9 ks for the FPMA and the FPMB data sets, respectively.

\subsection{Palomar}
An optical observation of the source, quasi-simultaneous to the {\it NuSTAR} pointing, was performed using the Double Spectrograph (DBSP) on the Hale 200$''$ Telescope at the Palomar Observatory on 2021 March 18. We obtained a single 300 s observation (PI: D. Stern) at the parallactic angle through the 1.5$''$ slit in photometric, good-seeing conditions. Data were flux calibrated (but not corrected for telluric absorption) using spectrophotometric standard stars.

\begin{figure}
\centering
  \epsfig{file=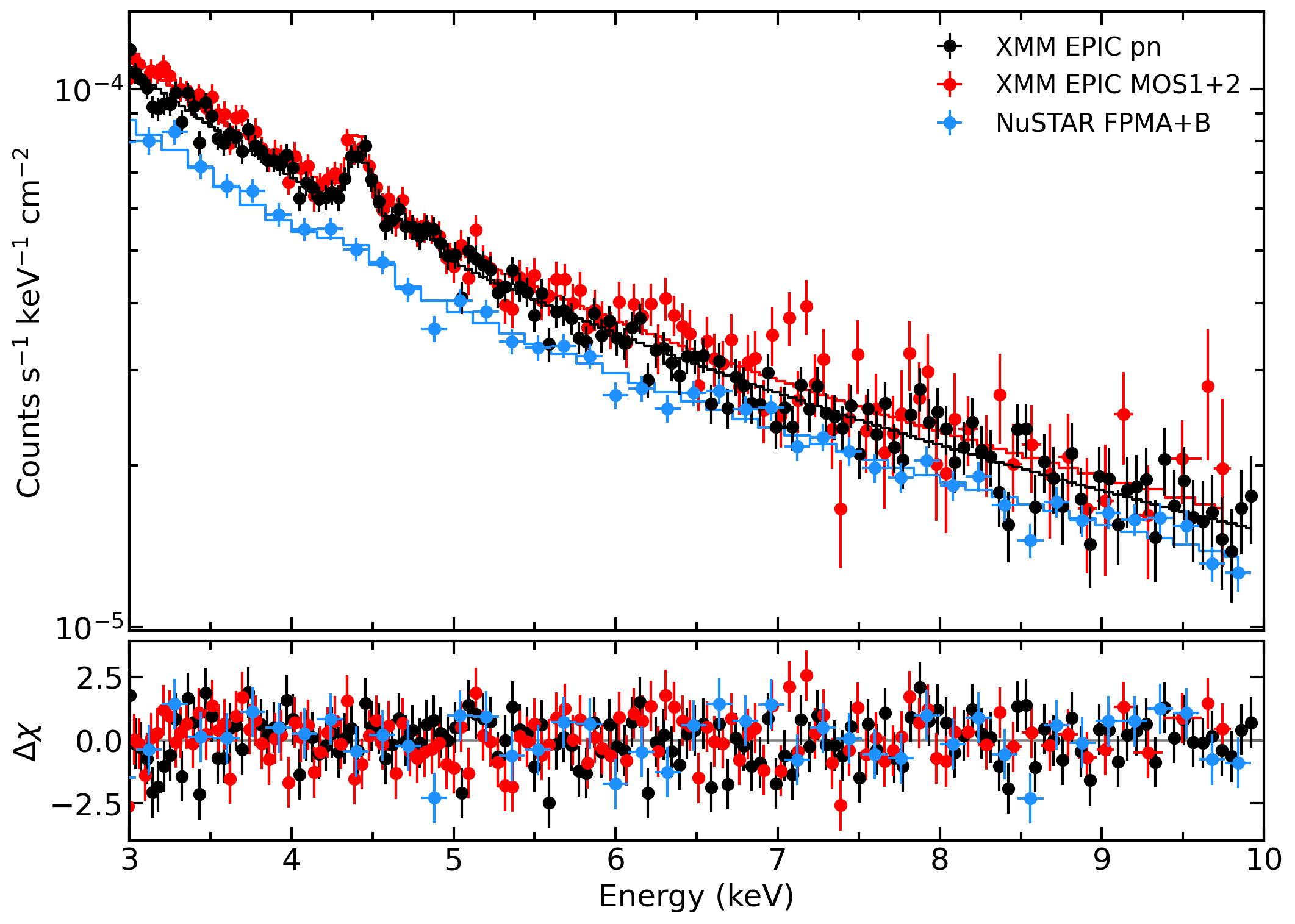, width=\columnwidth}
  \caption{Best fit in the 3-10 keV band. FPMA/B data are co-added, for the sake of visual clarity, using the {\sc setplot group} command in {\sc Xspec}.}
  \label{3_8keV}
\end{figure}

\section{X-ray spectroscopy}\label{sec:xray}
\subsection{3-10 keV data analysis}
\begin{figure}
\centering
  \epsfig{file=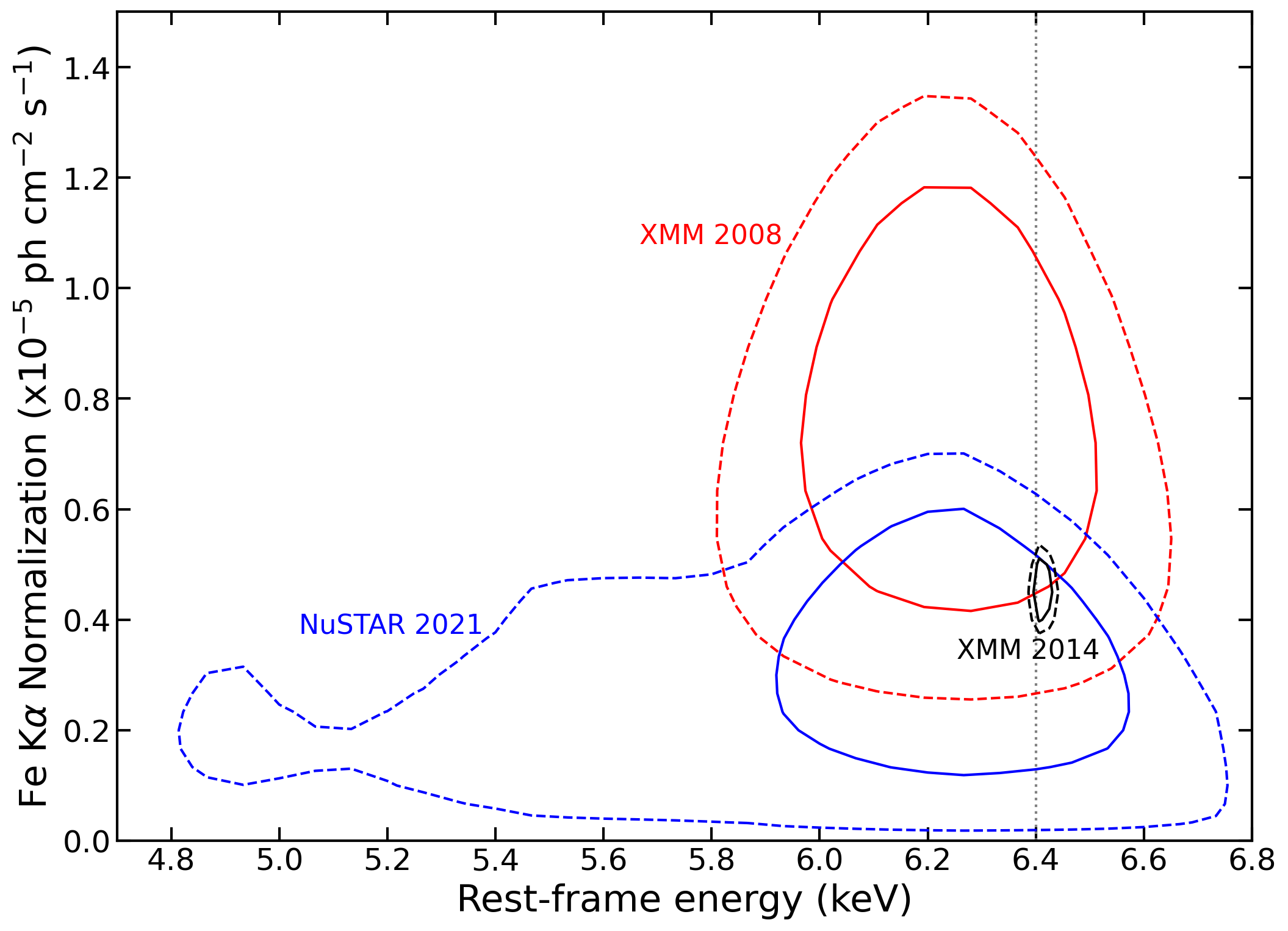, width=\columnwidth}
  \caption{Contour plots between the energy centroid and normalization of the Fe K$\alpha$ emission line, for the three epochs. Solid and dashed lines indicate 68\% and 90\% confidence level contours, respectively.}
  \label{cplot_feka}
\end{figure}

We start our analysis by fitting the pn, MOS12 and FPMA and FPMB spectra of the source in the 3-10 keV energy band. The model is composed of an absorbed redshifted power law ({\sc TBabs $\times$ zpow} in {\sc Xspec}) to reproduce the nuclear continuum, assuming a Galactic column density N$_{\rm H}=3.43\times10^{20}$ cm$^{-2}$ \citep{hi4pi}. Three multiplicative constants are included in the fit to take into account cross-calibration uncertainties between the pn and MOS12 detectors and flux variations between the 2014 and 2021 XMM and {\it NuSTAR} pointings. We obtain $\chi^2$/dof=469/357 and no statistically significant change of the photon index between the four spectra is found. Since large residuals can be seen in the observed 4-5.5 keV energy band we included three narrow redshifted Gaussian lines to reproduce the fluorescence Fe K$\alpha$ and K$\beta$ emission lines and the highly ionized Fe XXV/XXVI K$\alpha$ lines. The final best fit $\chi^2$/dof is 318/348 and best fitting values can be found in Table \ref{fit_3_10}. We show in Fig. \ref{3_8keV} the best fit model, data and residuals between 3 and 10 keV. While the first two emission lines at E$_1=6.41_{-0.01}^{+0.02}$ keV and E$_2=6.78_{-0.23}^{+0.08}$ keV can be readily identified as neutral Fe K$\alpha$ and Fe {\sc xxv} He-$\alpha$, the third one can be a blend of Fe {\sc xxvi} Ly-$\alpha$ and the fluorescence Fe K$\beta$ lines. The expected Fe K$\beta$/Fe K$\alpha$ (core only) ratio ranges from 0.155 to 0.16 \citep{mol03} and we find Fe K$\beta$/Fe K$\alpha$=$0.15^{+0.15}_{-0.11}$. Given the large uncertainties, a possible contamination from Fe XXVI Ly-$\alpha$ cannot be discarded. Leaving the width of the neutral Fe K$\alpha$ emission line free to vary does not lead to a statistically significant improvement of the fit ($\Delta\chi^2$/dof=-3/1) and an upper limit $\sigma<95$ eV is obtained. The best fitting values for the equivalent width of the Fe K$\alpha$ are EW=55$\pm$6 eV and EW=38$\pm$30 eV for the 2014 XMM and 2021 {\it NuSTAR} observations, respectively. Once the 3-10 keV best fitting model is applied to the 2008 pn snapshot, a good fit is obtained ($\chi^2$/dof=85/73) and an equivalent width EW=100$\pm60$ eV is retrieved, in agreement with \citep{klm10} The equivalent width of the emission lines is calculated via the {\sc eqw} command in {\sc Xspec}.
Since the uncertainties on the energy centroids and normalizations drive the total uncertainty on the EW value, we now show in Fig. \ref{cplot_feka} the contour plots between the Fe K$\alpha$ energy centroid and normalization, when these parameters are left free between the three epochs (solid and dashed lines indicate 68\% and 90\% confidence levels, respectively).

\begin{table}
\begin{center}
\begin{tabular}{ccc}
 {\bf Parameter}& {\bf Value} & {\bf Value}   \\
  & {\bf (XMM)} & {\bf (NuSTAR)}    \\

 \hline
  & \\
$\Gamma$ & \multicolumn{2}{c}{$1.66\pm0.02$}  \\
Normalization & \multicolumn{2}{c}{$(1.29\pm0.05)\times10^{-3}$}\\
C$_{\rm pn-MOS12}$ & \multicolumn{2}{c}{$1.061\pm0.014$}  \\
C$_{\rm pn-FPMA}$ & \multicolumn{2}{c}{$0.89\pm0.02$}  \\
C$_{\rm pn-FPMB}$ & \multicolumn{2}{c}{$0.94\pm0.02$}  \\
 & & \\
\multicolumn{3}{c}{\bf Fe K$\alpha$ line ($\Delta\chi^2$/dof=-137/-3)}  \\
E (keV) & \multicolumn{2}{c}{$6.41^{+0.02}_{-0.01}$}\\
N (ph cm$^{-2}$ s$^{-1}$) & $(4.5\pm 0.6)\times10^{-6}$ & $(3.3\pm 2.5)\times10^{-6}$\\
EW (eV) & $55\pm 6$ & $38\pm30$\\
 & & \\
\multicolumn{3}{c}{\bf Fe {\sc XXV} He-$\alpha$ line ($\Delta\chi^2$/dof=-9/-3)}  \\
E (keV) &  \multicolumn{2}{c}{$6.78_{-0.23}^{+0.08}$} \\
N (ph cm$^{-2}$ s$^{-1}$)& $(0.9\pm 0.6)\times10^{-6}$ & $<1.5\times10^{-6}$ \\
EW (eV) & $12\pm 8$ & $<18$\\
 & & \\
\multicolumn{3}{c}{\bf Fe K$\beta$ line ($\Delta\chi^2$/dof=-6/-3)}  \\
E (keV) &\multicolumn{2}{c}{$7.04\pm 0.08$}  \\
N (ph cm$^{-2}$ s$^{-1}$)& $(0.7\pm 0.5)\times10^{-6}$ &$<1.0\times10^{-6}$ \\
EW (eV) & $11\pm 8$ & $<13$ \\
\hline
\end{tabular}
\end{center}
\caption{\label{fit_3_10} Best fit parameters of the 3-10 keV fit of the joint XMM+{\it NuSTAR} data. The normalization of the nuclear power law is in ph cm$^{-2}$ s$^{-1}$ keV$^{-1}$ units at 1 keV. All reported values are in the rest-frame of the source and the statistical improvements are referred to the previous model.}
\end{table}

\subsection{0.3-79 keV data analysis}

\begin{figure}
\centering
  \epsfig{file=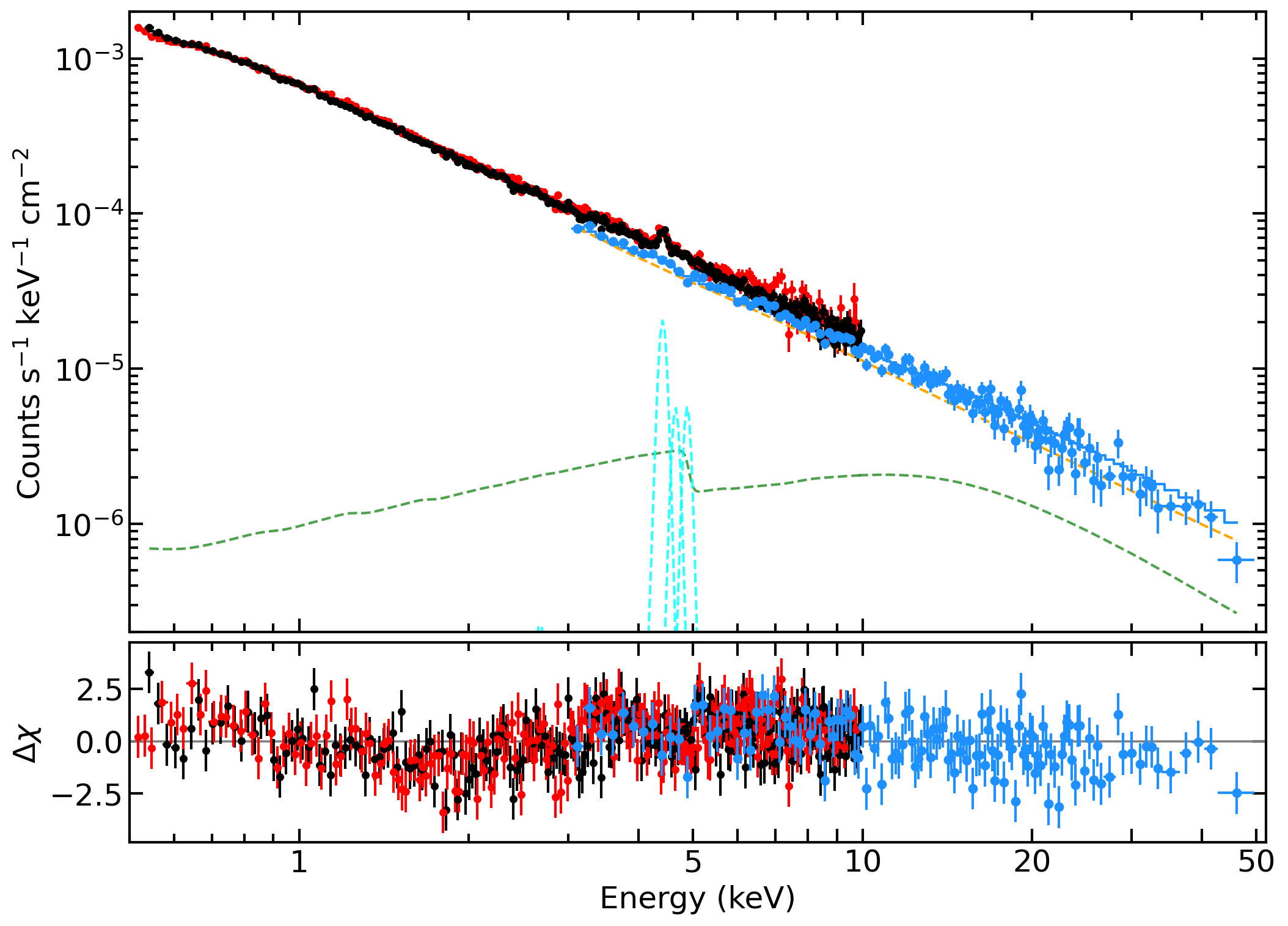, width=\columnwidth}
  \caption{Best fit model, data and residuals with {\sc pexrav} model, see text for details. Color coding for the different instruments follows the same scheme used for Fig. \ref{fit_3_10}. Clear residuals emerge at low energies.}
  \label{pexrav}
\end{figure}

We then extended the energy band including low and high energy data and re-fitted with the baseline model composed of an absorbed power law and three Gaussians. The resulting $\chi^2$/dof is good (1033/713=1.45) but some residuals can still be seen in the 3-10 keV band and at high energies.  We therefore included a Compton reflection component \citep[modeled with {\sc pexrav} in {\sc Xspec}:][]{mz95}, abundances were fixed to the solar values and the cosine of the inclination angle with respect to our line of sight was fixed to $\cos\theta=0.45$. The final $\chi^2$/dof is 908/712=1.27 and a Compton reflection fraction $R=0.55\pm0.10$ is inferred, with a photon index $\Gamma=1.77\pm0.01$. The best fit model and data are shown in Fig. \ref{pexrav}. Since some residuals are still present between 1 and 4 keV band we considered a toroidal model to reproduce the reflection component. 

\begin{figure}
\centering
  \epsfig{file=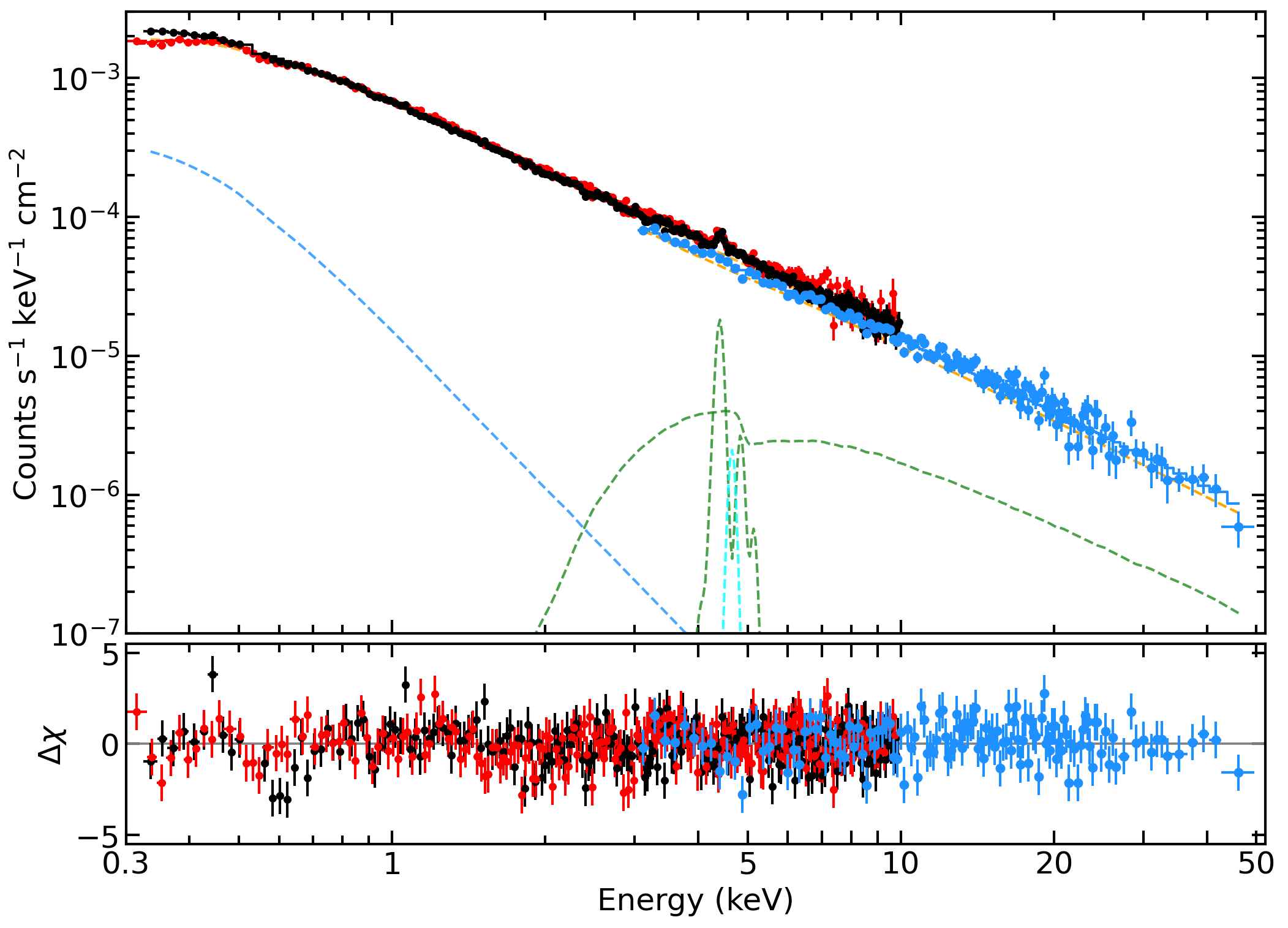, width=\columnwidth}
  \caption{Best fit model, data and residuals with the {\sc Borus} model, see text for details.  Color coding for the different instruments follows the same scheme used for Fig. \ref{fit_3_10}.}
  \label{bor}
\end{figure}

\begin{figure}
\centering
    \epsfig{file=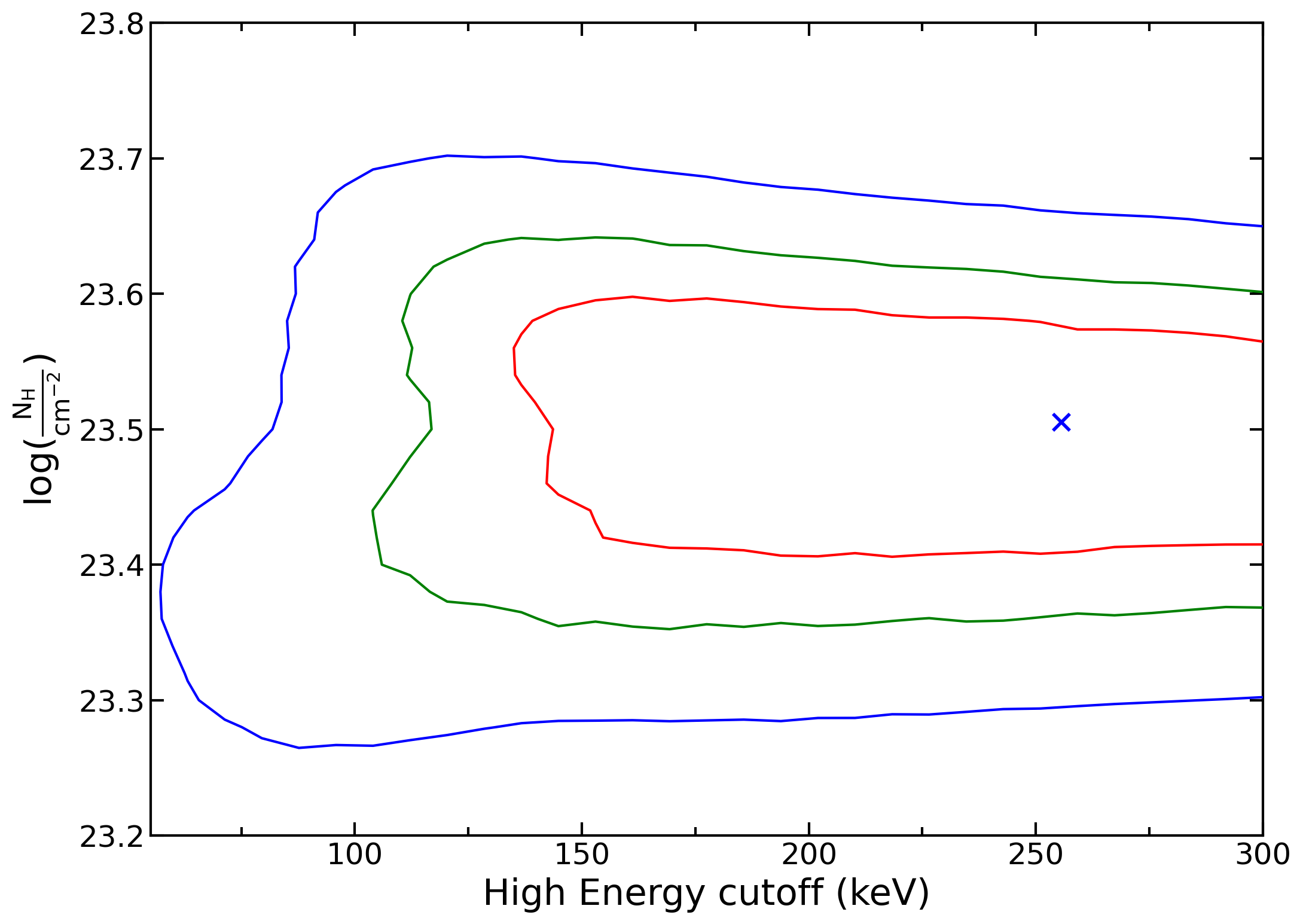, width=\columnwidth}
  \caption{Contour plots between the column density of the reflector N$_{\rm H}$ and the high energy cutoff E$_{\rm C}$ obtained with {\sc Borus} model. Red, green and blue solid lines indicate 68\%, 90\% and 99\% confidence levels, respectively.}
  \label{cplot} 
\end{figure}

We removed the two Gaussians accounting for the Fe K$\alpha$ and Fe K$\beta$ lines and {\sc pexrav} for the associated Compton reflection continuum and {\sc Xspec} tables generated with the Monte Carlo radiative transfer code {\sc Borus} \citep{bal18, bgc19} were included in the model. Different geometrical configurations for the circumnuclear material and physical assumptions on the input continuum can be tested. For the case of RBS 1055, we first consider two separate tables ({\sc borus02$_{-}$v170323k.fits} and {\sc borus02$_{-}$v170323l.fits}), one for the Compton reflection continuum and the other one for the associated emission lines. Furthermore, since no  evidence for neutral absorption in excess of the Galactic one was previously found \citep{klm10}, we assume two separate column densities for the absorber along the line of sight and for the scattering reflector. In particular, this is considered to be a sphere with conical cutouts at its poles, approximating a torus with variable covering factor. The half-opening angle of the polar cutouts, defined as ${\theta_{\rm tor}}$, is measured from the symmetry axis toward the equator, and ranges from 0$^{\circ}$ (which is a full covering sphere) to 83$^{\circ}$ (corresponding to a disk-like configuration covering 10\% of the solid angle).  In {\sc Xspec} the model reads as:
\begin{center}
 {\sc const$_{\rm cross-cal}\times$ TBabs$\times\Big($cutoffpl+zGauss+const$\times$Borus$_{\rm Scatt}$+\\+const$\times$Borus$_{\rm Lines}\Big)$ } .
 \end{center}
\noindent A constant covering factor of the torus C$_{\rm tor}$=$\cos({\theta_{\rm tor}})$=$0.5$ was assumed in this geometrical configuration. Normalizations, column densities, inclination angles and constants were linked between the two {\sc Borus} tables. Photon indices, cutoff energies and normalization were instead linked to the ones of the cutoff power law.  A best fit $\chi^2$/dof=719/715=1.01 is retrieved and best fitting parameters are reported in Table \ref{bestfitPar}.
Fig. \ref{bor} shows the best fit models, data and corresponding residuals. The contour plots between the column density of the reflector and the high energy cutoff E$_{\rm C}$ are shown in Fig. \ref{cplot}. 
A Compton-thin solution is preferred for the toroidal reprocessor and a column density N$_{\rm H}=(3.2_{-0.8}^{+0.9})\times10^{23}$ cm$^{-2}$ is measured. The primary continuum is well modelled with a power law component with $\Gamma=1.70^{+0.03}_{-0.05}$ and E$_{\rm C}>110$ keV. We note that the overall geometry of the torus is not self-consistent. The current configuration assumes $\theta_{\rm tor}$=60$^{\circ}$ and $\theta_{\rm inc}>70^{\circ}$, allowing our line sight to intercept part of the toroidal structure. However, we only retrieve an upper limit N$_{\rm H}<5\times10^{20}$ cm$^{-2}$ for the column density of an absorber covering the primary power law. If we impose  $\theta_{\rm inc}<\theta_{\rm tor}$ in the model a worse fit ($\Delta\chi^2=+14$) is obtained, with no statistically significant variation of best fitting parameters.\\

\begin{figure*}
\centering
    \epsfig{file=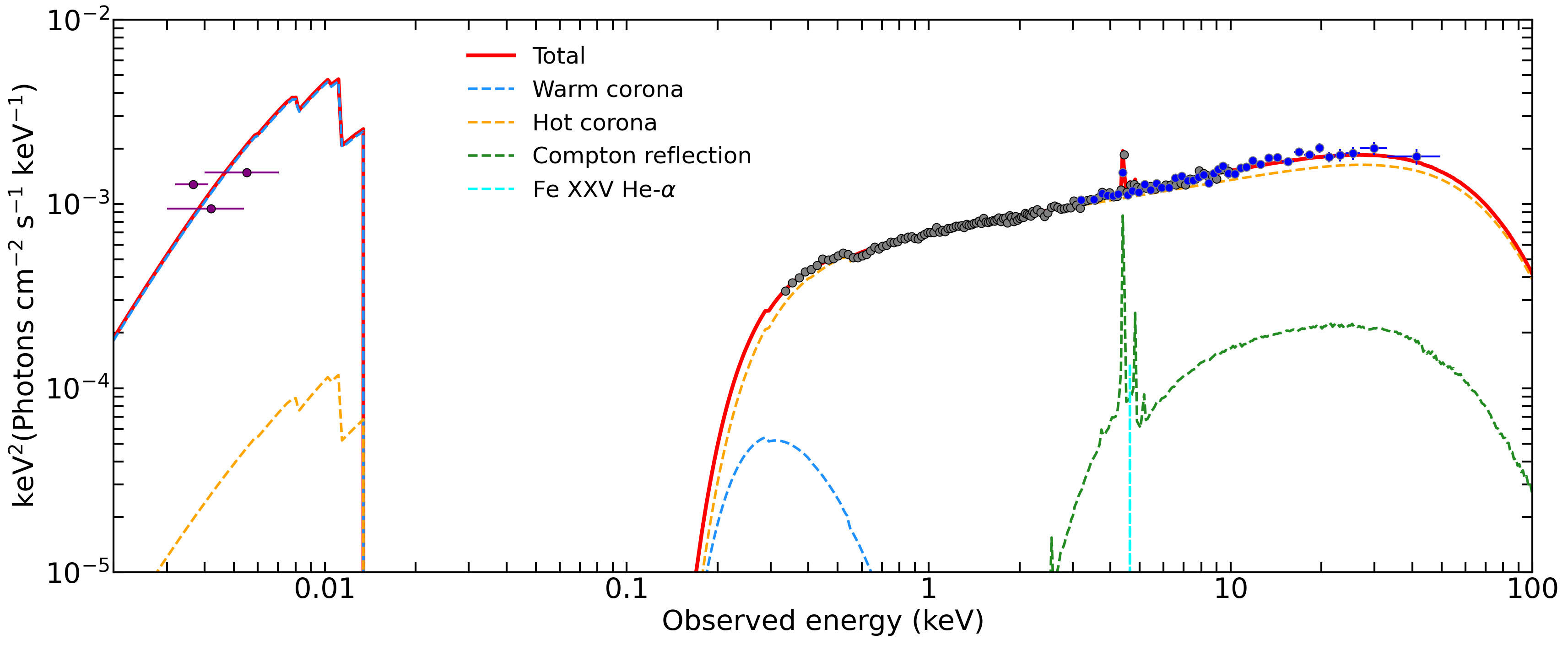, width=2.0\columnwidth}
  \caption{The total best fitting model is shown as a red solid line and UV/X-ray data are superimposed. The four spectral components are labeled and shown as dashed lines. XMM-{\it Newton} OM data are plotted in purple, pn data in grey and {\it NuSTAR} grouped FPMA/B data in blue. Data have been rebinned in energy for the sake of visual clarity.}
  \label{models}
\end{figure*}

\begin{figure}
\centering
    \epsfig{file=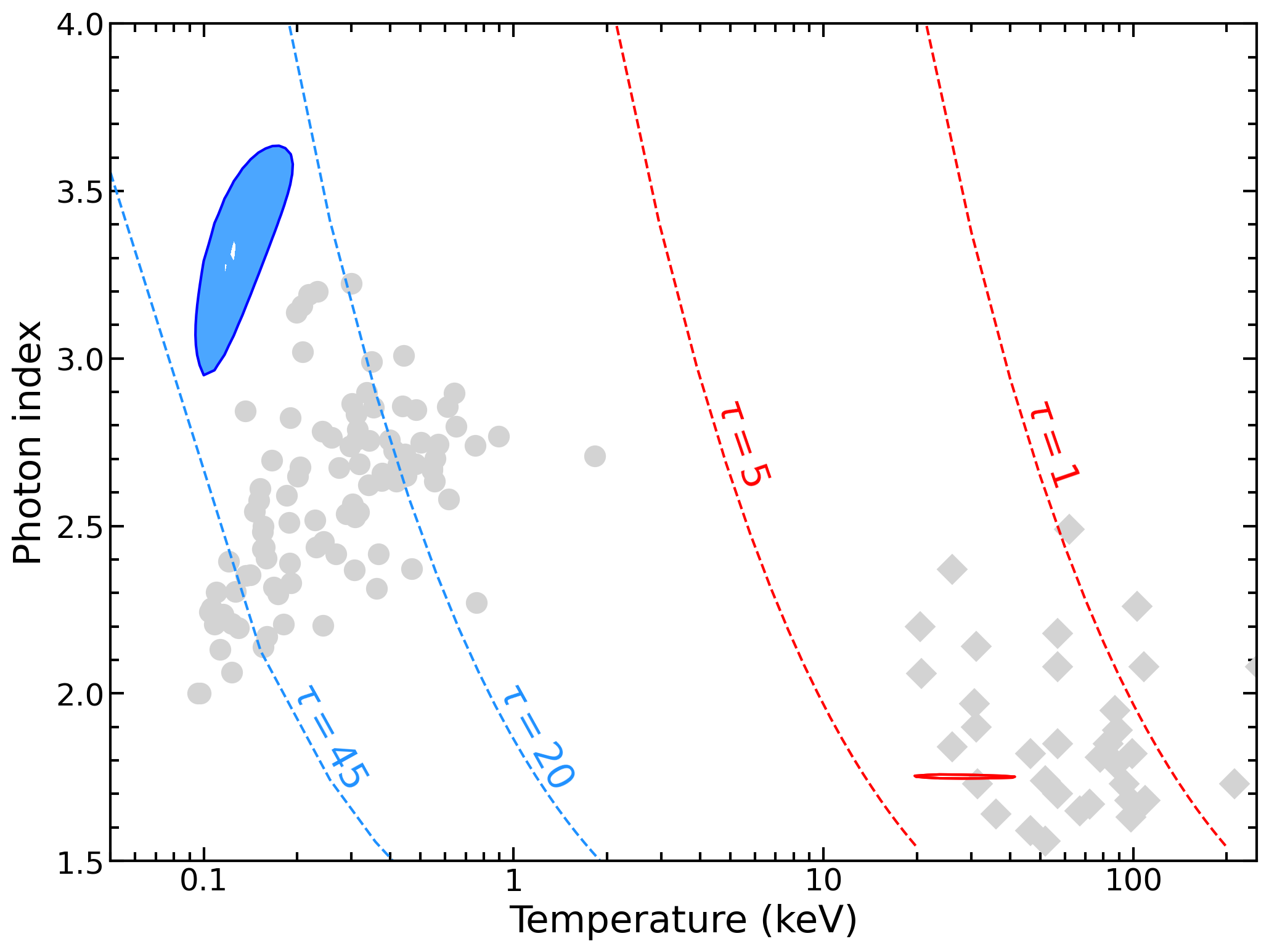, width=\columnwidth}

  \caption{Contour plots obtained with two {\sc Nthcomp} components for the warm and hot coronae are shown in blue and red, respectively. Shaded regions indicate 68\% confidence levels. Dashed red and blue lines indicate different Thomson optical depths obtained with the formula (A1) in \citet{zjm96}. Grey circles and diamonds are data points reported in \citet{pud18} for their sample of 22 sources.}
  \label{twocoronae} 
\end{figure}

Following \citet{klm10}, we checked for a soft excess component extending the energy range of the pn-MOS spectra down to 0.3 keV, therefore the final fit is between 0.3 and 79 keV. The resulting residuals were accounted for by an additional {\sc zpowerlaw} component in {\sc Xspec}, which largely improves the fit ($\Delta\chi^2$/dof=-67/-2), for a final $\chi^2$/dof=763/736=1.03. Some residuals are still present around $\sim$0.6 keV and the addition of an absorption line at E=0.88$\pm0.03$ keV with normalization N=$(-1.3\pm0.3)\times10^{-5}$ ph cm$^{-2}$ s$^{-1}$ improves the fit ($\Delta\chi^2$/dof=-21/-2). However, this absorption feature is detected in the EPIC-pn spectrum only and if a warm absorber component is included in the model ({\sc zxipcf} in {\sc Xspec}) only a marginal improvement of the overall fit is found ($\chi^2$/dof=751/734). The best fitting column density and ionization parameter are N$_{\rm H}<8\times10^{21}$ cm$^{-2}$ and $\log(\frac{\xi}{{\rm erg\ cm\ s{^{-1}}}})=3.0_{-1.1}^{+0.3}$.\\
\indent We then fixed the inclination angle, the photon index and the high energy cutoff of the reflection component to the best fitting ones, removed the soft and hard power laws and included the Comptonization model {\sc Nthcomp} \citep{zjm96,zds99}. The temperature of the blackbody component kT$_{\rm BB}$ is fixed to 5 eV, assuming a black hole mass M$_{\rm BH}=6.5\times10^8$ M$_{\odot}$. We have two different components, one for the soft excess and the other one for the hard power law continuum. We substituted the two {\sc Borus} tables with the {\sc borus12$_{-}$v190815a.fits} one, in which the intrinsic continuum is based on the {\sc Nthcomp} component and both the Compton scattered continuum and emission lines are included. The final {\sc Xspec} model reads as:
\begin{center}
 {\sc const$_{\rm cross-cal}\times$ TBabs$\times$\Big(Nthcomp$_{\rm W}$+Nthcomp$_{\rm H}$+zGauss+\\const$\times$Borus$_{\rm Nth}$\Big) } .
 \end{center}
A photon index $\Gamma^{\rm W}=3.30\pm0.37$ and a temperature kT$_e^{\rm W}=0.12_{-0.03}^{+0.08}$ keV  are inferred for the soft Comptonization component while $\Gamma^{\rm H}=1.75\pm0.01$ and kT$_e^{\rm H}=30_{-10}^{+40}$ keV are found for the hard one, with a resulting $\chi^2$/dof=765/740=1.03. These two pairs of values can be translated into optical depths $\tau$, using relation (1) in \citet{mpt19}, leading to Thomson optical depths $\tau_S=30_{-10}^{+15}$ and $\tau_H=3.0^{+1.0}_{-1.4}$ for the warm and hot coronal components, respectively. The final best fitting model is shown in Fig. \ref{models}, in which OM data points are also included. The three main components of the model can be clearly seen: while the warm (in blue) and hot (in orange) coronae account for the power-law like continua from the UV to the hard X-rays, the Compton thin reflector (in green) reproduces the Fe K$\alpha$ and its associated continuum. Fig. \ref{twocoronae} shows the contour plots between the photon indices and the temperatures obtained, using 68\% confidence levels. Grey data points are the best fitting values taken from \citet{pud18} for a sample of 22 radio quiet AGN.
\begin{table}
\begin{center}
\begin{tabular}{cc}
 {\bf Parameter}& {\bf Value} \\
 \hline
$\Gamma$ & $1.70_{-0.05}^{+0.03}$\\
E$_{\rm C}$ (keV) &  $>110$\\
N &$(1.39_{-0.01}^{+0.01})\times10^{-3}$ \\
$C$ & $1.2\pm0.2$ \\
N$_{\rm H}$ (cm$^{-2}$) & $(3.2_{-0.8}^{+0.9})\times10^{23}$\\
$\theta_{\rm obs}$ ($^{\circ}$) & $>70$ \\ 
$\Gamma_s$ & $3.64_{-1.04}^{+1.14}$\\
N$_s$ &$(8.2_{-4.5}^{+6.5})\times10^{-5}$ \\
C$_{\rm pn-MOS12}$ & $1.049\pm0.006$\\
C$_{\rm pn-FPMA}$ &  $0.89\pm0.02$\\
C$_{\rm pn-FPMB}$ & $0.94\pm0.02$\\
F$_{2-10\ \rm keV}$ & $(2.7\pm0.1)\times10^{-12}$ \\
L$_{2-10\ \rm keV}$ & $(2.0\pm0.2)\times10^{45}$ \\
$\chi^2/$dof &763/736 \\
\hline
\end{tabular}
\end{center}
\caption{\label{bestfitPar} Best fit parameters of the joint XMM+{\it NuSTAR} data analysis, obtained with the {\sc Borus} component for the Compton reflection. Normalizations are in ph cm$^{-2}$ s$^{-1}$ keV$^{-1}$ units at 1 keV, fluxes and luminosities are in erg cm$^{-2}$ s$^{-1}$ and erg s$^{-1}$ units, respectively. $C$ indicates the multiplicative constant for the {\sc Borus} tables while C$_{\rm pn-MOS12}$, C$_{\rm pn-FPMA}$ and C$_{\rm pn-FPMB}$ are the three cross-calibration factors which multiply the total model.}                
\end{table}

\section{Optical spectroscopy}
\subsection{Data analysis}
\begin{figure}[h!]
\centering
  \epsfig{file=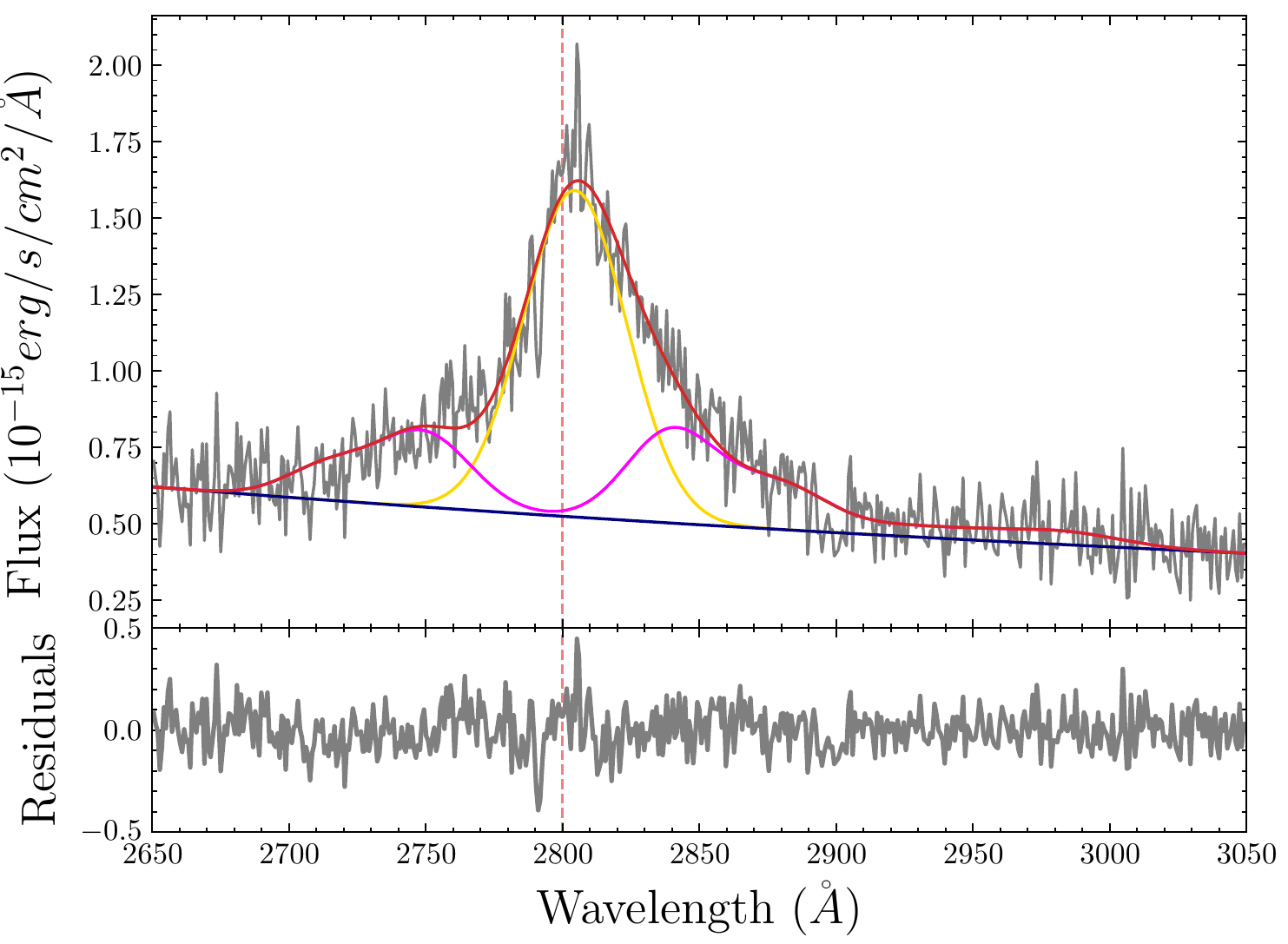, width=1\columnwidth}
  \epsfig{file=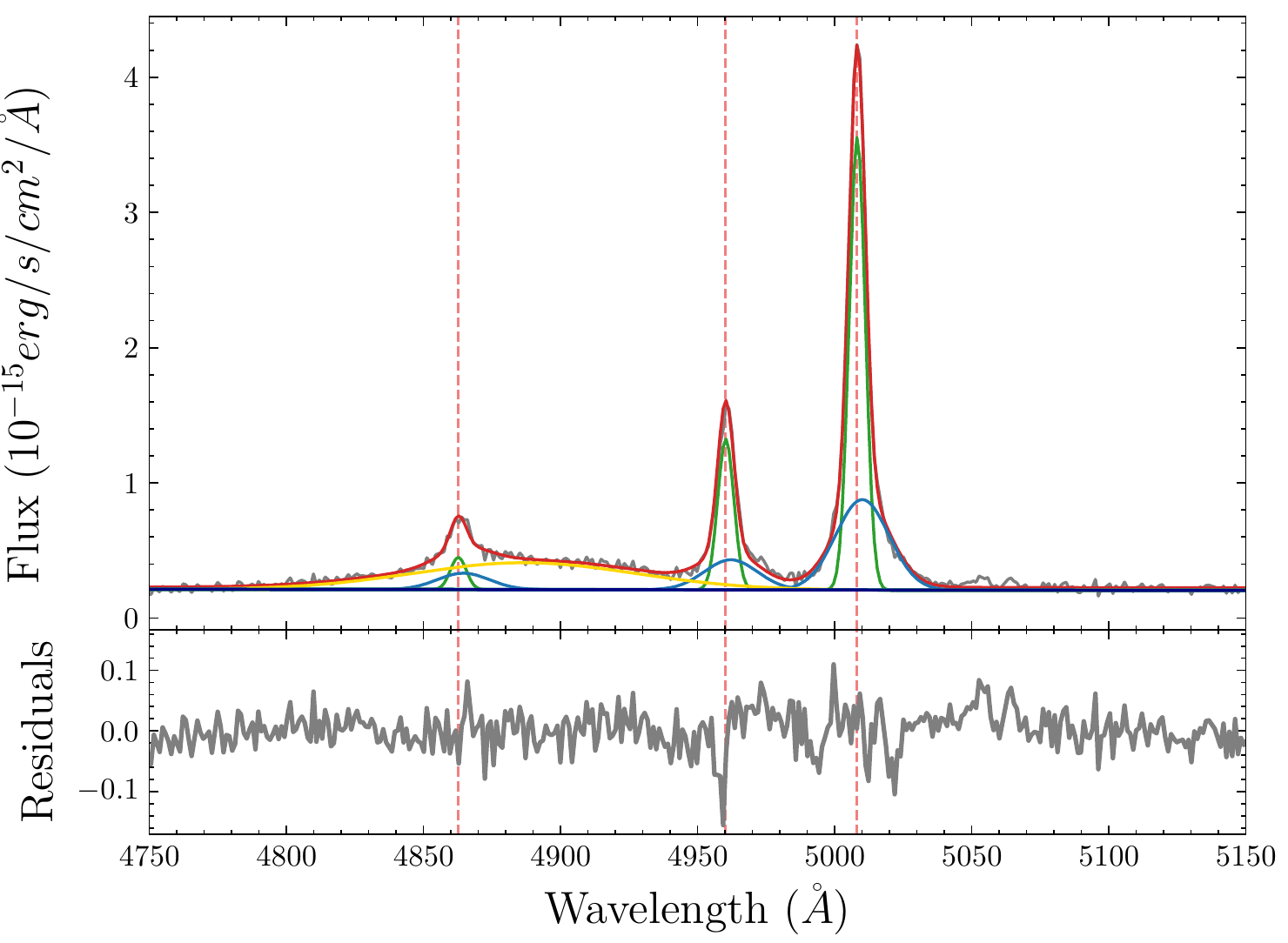, width=1\columnwidth}
  \epsfig{file=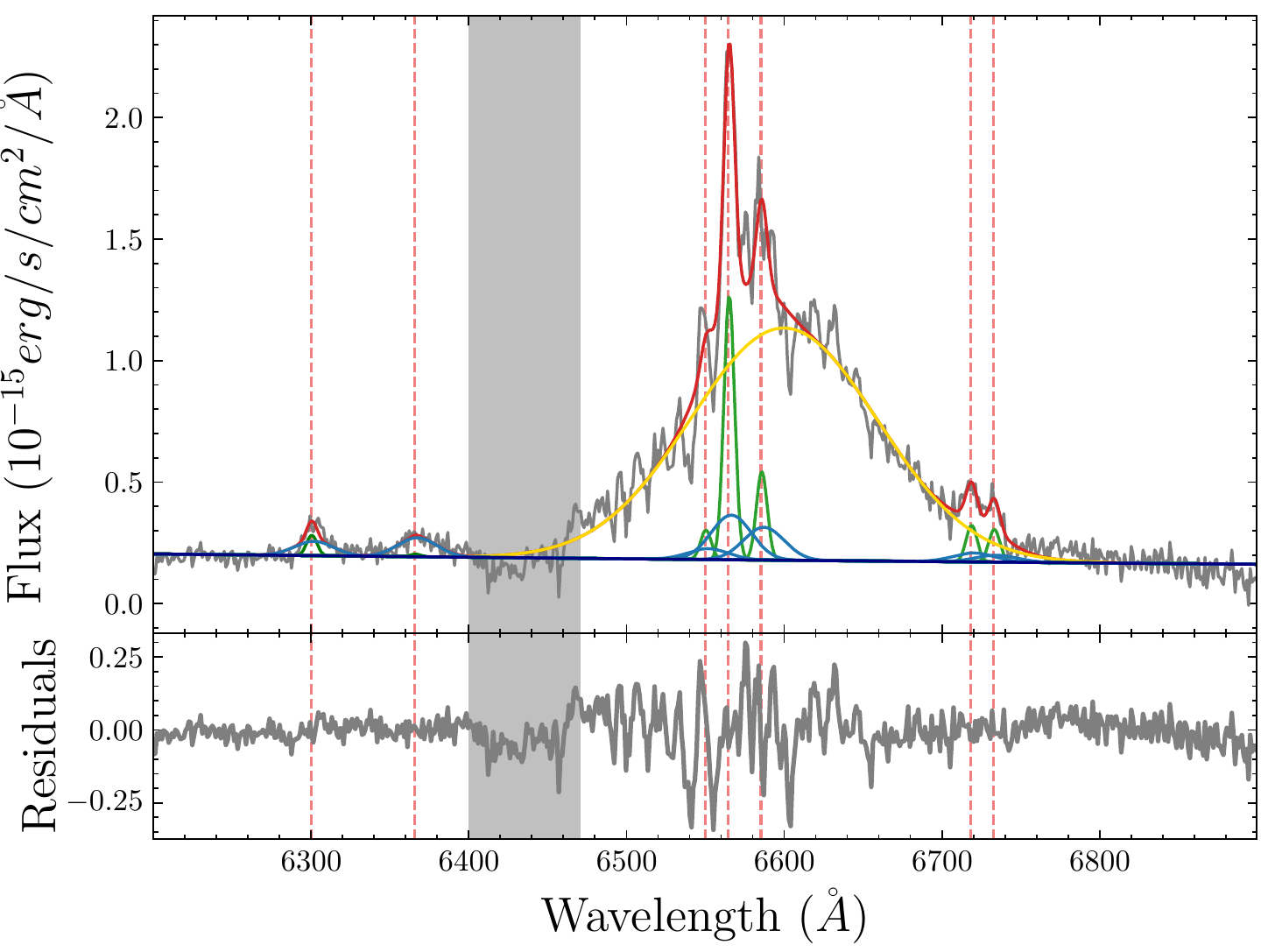, width=0.98\columnwidth}
  \caption{Optical spectroscopy. (a) The MgII line spectrum, (b) H$\beta$-[OIII] region and (c) H$\alpha$-[SII] region. The red curves represent the best-fit models. Gold Gaussian components represent the emission originating from the BLR, the magenta curve refers to the iron emission and navy blue power-law curves represent the continuum emission. Green Gaussian components reproduce the NLR emission and blue ones trace the outflowing gas in the NLR. Dashed vertical red lines indicate the location of the emission lines of interest at the systemic redshift. The vertical grey shaded region marks the channels affected by telluric absorptions which were masked during the fitting procedure.}
  \label{fig:opt_f}
\end{figure}

The red and blue channels of DBSP cover the MgII, H$\beta$, [OIII] doublet
and the H$\alpha$ emission lines. To model the line profiles, we fitted separately the emission lines as shown in Fig. \ref{fig:opt_f}, with one or two Gaussian components with a local continuum characterization, described by a powerlaw function. Specifically, (a) the MgII line is modelled with a broad Gaussian component to describe the emission from the broad line region (BLR), and the adjacent wavelength range, contaminated by FeII multiplets, with FeII templates from \cite{Popovic2019}, which are convolved with a Gaussian function with a full-width half maximum (FWHM) in the range 1000-5000 km s$^{-1}$; (b) H$\beta$ and [OIII] doublet lines with one Gaussian component to reproduce the systemic emission from the narrow line region (NLR) and an additional Gaussian function for any possible blueshifted/asymmetric emission associated with an outflowing gas. Moreover, we added a broad Gaussian component for the H$\beta$ line to reproduce the emission from the BLR; (c) similarly, the H$\alpha$, [NII]6548,6583 \AA\ emission line doublet and [SII] doublet are modelled with two Gaussian components, systemic plus any contribution from the outflowing gas, and a broad Gaussian component for H$\alpha$ originating in the BLR. 

The centroids and the velocity dispersion of similar components are tied together. Further, the outflow components of the H$\alpha$+[NII]+[SII] system have the same velocity dispersion as the [OIII] outflow component, to avoid degeneracy. [NII]6548 \AA\ /[NII]6583 \AA\ and [OIII]5007 \AA\ /[OIII]4959 \AA\ flux ratios were fixed to 3 based on the atomic transition probability (\citealt{Osterbrock2006}). We modelled the FeII emission with the observational templates of  \cite{Boroson1992}, \cite{Veron2004}, and \cite{Tsuzuki2006}, however we do not detect a strong FeII component in the region of H$\beta$.  The best-fit model is chosen with $\chi^2$ minimization process. The best-fitting parameters are presented in Table \ref{tab:opt}, and the best-fit models in Fig. \ref{fig:opt_f}. The uncertainties of free parameters are calculated using a Monte-Carlo approach: we added random Gaussian noise to the best-fit model with dispersion equal to the rms of the spectrum and repeated the fit 1000 times. The associated errors are estimated using the 84$^{th}$ and 16$^{th}$ percentiles of the parameter distribution.

\subsection{Continuum and emission line luminosity}\label{sec:continuum}
From the emission line modelling, we derived a more accurate estimate of the redshift by adopting as systemic line the [OIII]$\lambda$5007 narrow component ($z$=0.452$\pm$0.001). We found that the H$\beta$ BLR line centroid is redshifted with respect to the expected line position (4862 \AA) of $\Delta \rm v_{H\beta}$=1500$\pm$100 km s$^{-1}$ and the BLR H$\beta$ line luminosity is $L\rm^{BLR}_{H{\beta}}$=1.56$\pm$0.07$\times 10^{43}$ erg s$^{-1}$. Similarly, we obtained BLR H$\alpha$ line luminosity $L\rm^{BLR}_{H{\alpha}}$=1.11$\pm$0.03$\times\  10^{44}$ erg s$^{-1}$ and consistent redshifted centroid value as that found for the BLR component of H$\beta$ ($\Delta \rm v_{H\alpha}$=1600$\pm$100 km/s). We computed the observed Balmer decrement, defined as the line ratios of BLR H$\alpha$/H$\beta$ and found a value of 7.08$\pm$0.18. 

\cite{Baron2016} analysing a sample of $\sim$5000 Type 1 AGN at z$\sim$0.4 found a mean H$\alpha$/H$\beta$ $\sim$3 with a broad distribution in the range 1.5–4, suggesting that Case B recombination cannot explain the ratio in all sources. We calculated the optical reddening as E(B-V)=1.97$\times$log10((H$\rm\alpha$/H$\rm\beta$)/(H$\rm\alpha$/H$\rm\beta$)$\rm_{intrinsic}$) (\citealt{Osterbrock1989}), where the intrinsic Balmer decrement is assumed to be 3 (\citealt{Baron2016}) and derived E(B-V)=0.73. However, by comparing the optical spectrum of our target with that of the composite AGN spectrum of \cite{VandenBerk2001}, we found no extinction for the continuum, confirming that our target actually has a blue optical continuum (see Sec. \ref{sec:discussion} for a discussion).


For the NLR we found H$\alpha$/H$\beta$=5.4$\pm$0.4, lower than the BLR value but still larger than the intrinsic Balmer ratio for a low-density environment as the NLR, 2.74–2.86, assuming a case B recombination (\citealt{Osterbrock2006}). 
The lower value found for the NLR suggests a different level of reddening in the two regions and may implies that the extinction of the BLR is due to dusty structures that are more compact than the NLR.

\subsection{BH mass and Eddington ratio}
Single-epoch BH masses are usually estimated for Type 1 AGN from the broad emission lines (e.g. \citealt{Greene2007}, \citealt{Vestergaard2009}). Assuming that the BLR clouds are in virial motion, the BH mass is proportional to RV$^2$/G, with R the radius of the BLR and V the gas velocity. These two parameters can be estimated from the width of the broad emission lines and the radius from the radius-luminosity relation defined by reverberation mapping (e.g. \citealt{Bentz2013}). From the best-fitting results we calculated the monochromatic luminosity at 3000 \AA\ ($L\rm_{3000}$=9.8$\pm$0.16$\times$10$^{44}$ erg s$^{-1}$), 5100 \AA\ ($L\rm_{5100}$=8.10$\pm$0.06$\times$10$^{44}$ erg s$^{-1}$) and H$\alpha$ BLR line luminosity (see Table \ref{tab:BH}), as well as the FWHM of H$\beta$, H$\alpha$ and MgII lines to derive the BH mass estimated by using the calibration of \citet{Bongiorno2014} for H$\beta$ and MgII and \citet{Greene2005} for H$\alpha$. We found the BH mass in the range 2.79$\times$10$^{8}$-1.19$\times$10$^{9}$ M$\rm\odot$, according to different BLR emission lines, with an average value of <$M\rm_{BH}$>=6.52$\times$10$^{8}$ M$\rm\odot$, which we used as our preferred value hereafter.

We estimated the bolometric luminosity of the AGN assuming a bolometric correction factor for the continuum luminosity $\lambda$5100$L\rm_{5100}$, adopting the prescription proposed by Saccheo et al (2022), in prep (K$_{5100 \AA}$ $\sim$4.80$\pm$1.54), and found $L\rm_{Bol}$= 3.86$\times$10$^{45}$ erg s$^{-1}$. From the BH mass and bolometric luminosity it is possible to estimate the Eddington ratio defined as $\lambda\rm_{Edd}$= L$\rm_{Bol}$/L$\rm_{Edd}$, and we found
$\lambda\rm_{Edd}$=0.05, by assuming <$M\rm_{BH}$>.

\subsection{Outflow properties}\label{sec:outflow}

We traced the outflow in the NLR in its warm ionized phase using the [OIII]5007 emission line, since it is sensitive to the gas temperature and ionization parameter. Previous works alternatively used H$\beta$ line (e.g. \citealt{Liu2013}; \citealt{Harrison2014}); however, in our case, the H$\beta$ line is mostly dominated by the component arising from the BLR gas, therefore the [OIII] is the preferred outflow tracer. 

We estimated the outflowing gas mass $M\rm_{out}$ and mass outflow rate $\dot{M}\rm_{out}$ following the method presented in \cite{bpv17}, assuming a spherically/biconically symmetric mass-conserving free wind with a mass outflow rate and velocity that are independent of radius (\citealt{Rupke2002,Rupke2005}), and that most of the oxygen consists of [OIII] ions.
We used the relation by \cite{Carniani2015} and derive the outflowing ionised gas mass from the luminosity associated with the broad [OIII] emission:

\begin{eqnarray}
\log\frac{M{\rm_{out}}}{M_{\odot}} = 7.6 +{\rm\log\Big( \frac{C}{10^{[O/H]-[O/H]_{\odot}}}\Big)} + \log\Big( \frac{L{\rm_{[OIII]}^{out}}}{10^{44}\ {\rm erg\ s} ^{-1}}\Big)+ \nonumber \\ 
- \log  \Big(\frac{<n{\rm_e}>}{10^3\ {\rm cm}^{-3}}\Big) \hspace{5.5 cm} 
\end{eqnarray}

where C = <$n\rm_e$>$^2$/<$n^2\rm_e$>, [O/H]-[O/H]$_{\odot}$ is the gas metallicity relative to the solar value, $L\rm^{out}_{[OIII]}$ is the outflowing [O III]5007  luminosity and <n$_e$> is the average electron density. This last parameter can be estimated from the line ratio of [SII]$\lambda$6716,6731 doublet (\citealt{Peterson1997}) of the outflowing component. Assuming an electron temperature of $\sim 10{^4}$ K, a typical value for AGN outflows (e.g. \citealt{Perna2017} and references therein), we found a value of $n\rm_e\cong$ 177 cm$^{-3}$. Assuming a solar metallicity and C $\approx$ 1, we derived $M\rm_{out} = 2.84 \times 10^7$ M$_{\odot}$. We note that this measurement strongly depends on the electron density. We also derived the mass outflow rate  $\dot{M}\rm_{out}$ from the fluid-field continuity equation for a local estimate of the mass rate at a given radius, as follows:
\begin{equation}
    \dot{M}{\rm_{out}} = 3\frac{M{\rm_{out}} \rm{v_{max}}}{R{\rm_{out}}}
\end{equation}

where v$\rm_{max}$ is the outflow velocity defined as |$\Delta$v$\rm|^{out}_{[OIII]}$ + 2$\sigma\rm_{[OIII]}^{out}$ ([OIII]
|$\Delta$v| is the velocity shift between narrow and outflow [OIII] emission centroids and $\sigma$ is the velocity dispersion of the [OIII] outflow component) and $R\rm_{out}$ is the spatial extent of the outflow. Since we do not have spatial information, we assumed $R\rm_{out}$=4.4 kpc, equal to the half-slit width. From the best-fitting model parameters, we derived the velocity shift and velocity dispersion of the outflow component, and therefore v$\rm_{max}$, as reported in Table \ref{tab:opt}. In this way we obtained $\dot{M}\rm_{out}\sim$ 25 M$_{\odot}$ yr$^{-1}$.  
From $\dot{M}\rm_{out}$ we then derived the kinetic power associated with the outflow, $\dot{E}\rm_{kin}$= $ \frac{1}{2} \dot{M}\rm_{out}$v$\rm_{max}^2$, and find a value of 1.3$\times$10$^{43}$ erg s$^{-1}$, corresponding to a coupling efficiency with the interstellar medium $\dot{E}\rm_{kin}$/$L\rm_{Bol}$ $\approx$ 0.33\%. Theoretical models predict a coupling of 0.1-5\% for AGN-driven outflows (\cite{King2005}). The value found for RBS1055 might be sufficient for an efficient feedback mechanism which will affect the gas content and star formation rate in the host galaxy. However, IFU spectroscopy is necessary to accurately determine the exact extension of the outflow and the spatial distribution of the density of the outflowing gas. We note that, given the typical multi-phase nature of AGN-driven outflows (\citealt{Cicone2018}; \citealt{Bischetti2019}; \citealt{Fluetsch2021}), the derived $\dot{E}\rm_{kin}$ should be considered as a fraction of the total kinetic power of the outflow in RBS1055 including all gas (i.e. molecular and atomic) phases.

\begin{table*}
\begin{center}
\begin{tabular}{cccccccc}

 \hline
\bf{$FWHM\rm_{[OIII]}^{Narrow}$} & \bf{$L\rm_{[OIII]}^{Narrow}$} & \bf{$FWHM\rm_{[OIII]}^{Broad}$} & \bf{$L\rm_{[OIII]}^{Broad}$} & v$\rm_{max}$ & $M\rm_{out}$&  $\dot{M}\rm_{out}$ & $\dot{E}\rm_{kin}$  \\
km s$^{-1}$ & 10$^{43}$ erg s$^{-1}$ & km s$^{-1}$ & 10$^{43}$ erg s$^{-1}$ & km s$^{-1}$ & 10$^7 M\rm \odot$ &  $M\rm \odot$ yr$^{-1}$ & 10$^{43}$ erg s$^{-1}$  \\
(1) & (2) & (3) & (4) & (5) & (6) & (7) & (8) \\

\hline
420$\pm$10 & 1.91$\pm$0.03 & 1370$\pm$50 & 1.26$\pm$0.03 & 1280$\pm$50 & 2.84$\pm$0.12 & 25.4$\pm$1.5 & 1.3$\pm$0.12 \\
\hline

\end{tabular}
\end{center}
\caption{[OIII]$\lambda$5007 emission line properties.}\label{tab:opt}
Note.  (1) FWHM of the narrow component, (2) [OIII] luminosity  of the narrow component, (3) FWHM of the broad component, (4) [OIII] luminosity  of the broad component, (5) maximum velocity as defined in Sec. \ref{sec:outflow}, (6) mass of the outflowing gas, (7) mass outflow rate and (8) outflow  kinetic power.
\end{table*}

\begin{table*}
\begin{center}
\begin{tabular}{cccccccccc}

 \hline
\bf{$FWHM\rm_{MgII}$} & \bf{$L\rm_{MgII}$} & \bf{$FWHM\rm_{H\beta}$} & \bf{$L\rm_{H\beta}$} & \bf{$FWHM\rm_{H\alpha}$} & \bf{$L\rm_{H\alpha}$} & $M\rm_{BH}^{MgII}$ & $M\rm_{BH}^{H\beta}$ & $M\rm_{BH}^{H\alpha}$ & $\lambda_{Edd}$ \\
km s$^{-1}$ & 10$^{43}$ erg s$^{-1}$ & km s$^{-1}$ & 10$^{43}$ erg s$^{-1}$ & km s$^{-1}$ & 10$^{43}$ erg s$^{-1}$ & M$_{\odot}$& M$_{\odot}$& M$_{\odot}$ \\

\hline
4730$\pm$180&3.87$\pm$0.17 & 5850 $\pm$200 & 1.56 $\pm$0.07 & 6340$\pm$100 & 10.94$\pm$0.08 &2.79$\times$10$^{8}$ & 4.89$\times$10$^{8}$ & 1.19$\times$10$^{9}$ & 0.05 \\
\hline

\end{tabular}
\end{center}
\caption{BLR properties from MgII, H$\beta$ and H$\alpha$ emission lines.}\label{tab:BH} 
\textbf{Note}. We report the BH mass estimates from all the BLR emission lines, while for the Eddington estimate we report the value derived from the average BH mass <$M\rm_{BH}$>, which we assumed as our fiducial value.
\end{table*}

\section{Discussion}\label{sec:discussion}
We reported, in the previous sections, on the long observations of the bright quasar RBS 1055 with XMM-{\it Newton} in 2014 and with {\it NuSTAR} in 2021. A $\sim 10\%$ drop in the 2-10 keV flux (1.38-6.88 keV at the rest-frame of the source) is observed after seven years, from F$_{2-10}=(2.7\pm0.1)\times10^{-12}$ erg cm$^{-2}$ s$^{-1}$ to F$_{2-10}=(2.4\pm0.2)\times10^{-12}$ erg cm$^{-2}$ s$^{-1}$. At the redshift of the source, these fluxes correspond to L$_{2-10}=(2.0\pm0.2)\times10^{45}$ erg s$^{-1}$ and L$_{2-10}=(1.8\pm0.2)\times10^{45}$ erg s$^{-1}$, respectively. Assuming a black hole mass estimate of M$_{\rm BH}=6.5\times10^8$ M$_{\odot}$ (see Sect. 4.3) and adopting the bolometric corrections $K_X(L_X)$ from \citet{dbr20} (Equation 3) we retrieve bolometric luminosities L$_{\rm Bol}$=7.4$\times10^{46}$ erg s$^{-1}$ and 6.4$\times10^{46}$ erg s$^{-1}$ for the 2014 and 2021 epochs, respectively. The corresponding accretion rates are $\lambda_{\rm Edd}$=0.9 and $\lambda_{\rm Edd}$=0.8. The 2021 bolometric luminosity calculated from the 2-10 keV luminosity is a factor $\sim15$ higher than the one estimated from the continuum luminosity at 5100\AA. An important proxy of the interaction between the accretion disk and the corona is the slope of the power law connecting the rest-frame X-ray luminosity at 2 keV and the rest-frame UV luminosity at 2500 $\AA$ i.e., $\alpha_{\rm ox}=0.384\log({\rm L_{2500\ \AA}}/L_{\rm 2\ keV})$ (\citealt{tana79, steffen06, just07};\citealt{lusso10, lr17, rl19}). Interpolating the UV luminosities inferred with the UVW1 (2910 $\AA$) and U (3440 $\AA$) filters, we obtain a monocromathic 2500 $\AA$ luminosity $\log(\frac{{\rm L_{2500\ \AA}}}{\rm erg\ s^{-1}\ Hz^{-1}})=30.05$ and, using the best fit discussed in Sect. 3.2, a 2 keV luminosity $\log(\frac{{\rm L_{\rm 2\ keV}}}{\rm erg\ s^{-1}\ Hz^{-1}})=27.29$. We therefore infer an $\alpha_{\rm ox}=1.06$. This value lies at the very low end of the $\alpha_{\rm ox}$ distribution obtained from high-redshift quasars samples. If compared with the WISSH (WISE-SDSS selected hyper-luminous) quasars \citep[$z\sim$1.8-4.8][]{bpv17}, using the best fitting relation (3) in \citet{mpz17}, we obtain a difference $\Delta\alpha_{\rm ox}=0.39$. From Eq. (8) in \citet{vat13}, in which {\it Swift} observations of sources at $z\sim$0.01-0.4 are considered, we find $\Delta\alpha_{\rm ox}=0.35$. Applying the L$_{X}$-L$_{\rm UV}$ relation reported in \citet{rl19} to the measured monocromathic 2500 $\AA$ luminosity a $\log(\frac{{\rm L_{\rm 2\ keV}}}{\rm erg\ s^{-1}\ Hz^{-1}})\simeq26.2$ is retrieved, well below the $\log(\frac{{\rm L_{\rm 2\ keV}}}{\rm erg\ s^{-1}\ Hz^{-1}})=27.29$ observed value. This difference is in agreement with the one between the bolometric luminosities discussed above. We therefore conclude that RBS 1055, despite a $\sim10\%$ decrease in the total 2-10 keV flux in 2021, continues to have an extremely X-ray bright SED, 10-15 times higher than other objects in this redshift range.\\
\indent The detection of Compton reflection features are very scarce and poorly constrained in luminous AGN since they are typically derived from 0.5-10 keV spectra. They seem to suggest low values of the Compton reflection fraction $R$, i.e. much lower than unity \citep{rt00,page05}, even when {\it NuSTAR} observations are taken into account \citep{zcc18}. An anti-correlation exists between the Equivalent Width (EW) of the narrow neutral Fe K$\alpha$ line and L$_{2-10\rm\ keV}$ \citep[e.g. the IT effect;][]{it93,page04,bianchi07}. The IT effect is interpreted in terms of a decreasing covering factor of the Compton-thick toroidal reflector as a function of the increasing L$_{2-10}$. However, the quality of the data for objects with L$_{2-10}>10^{45}$ erg s$^{-1}$ is typically too poor even to put tight constraints on the EW of the Fe K$\alpha$ line \citep{jpg05}. RBS 1055 is a clear outlier of this observed correlation: considering the best fitting EW-L$_{2-10\rm\  keV}$ relation in \citet{bianchi07} for radio-quiet sources, we measured an Fe K$\alpha$ EW which is $\sim3\sigma$ above the EW=32 eV predicted value. However, this difference reduces to $\sim2\sigma$ if their EW-$\lambda_{\rm Edd}$ relation is considered. This suggests that, despite the high luminosity, the circumnuclear reflector still covers a large solid angle and it is found to be Compton-thin (with an equatorial column density N$_{\rm H}=3.2^{+0.9}_{-0.8}\times10^{23}$ cm$^{-2}$). Our best fit model, using {\sc Borus}, does not constrain the covering factor of the torus since we only retrieve $\theta_{\rm tor}>50^{\circ}$ when this parameter is left free (for an improvement of the fit $\Delta\chi^2$=-7 with one additional degree of freedom). This suggests that the circumnuclear torus in RBS 1055 is likely to be clumpy, as already observed in some local sources \citep{uru21, mbm16}. Even though only Compton-thick candidates were considered, according to recent {\it NuSTAR} samples \citep{bbs15, maz19}, this is the first tentative measurement of a torus covering factor above 10$^{45}$ erg s$^{-1}$. \\
\indent In the last few years, high quality {\it Chandra} data have shown that the Fe K$\alpha$ can be spatially extended both in Compton-thin \citep{yi21} and in Compton-thick galaxies \citep{mrw12, fabbiano17, jones21}. The projected distances range from tens of parsecs \citep[as in  the Circinus galaxy:][]{mmb13} up to the kpc scale \citep[as in NGC 1068:][]{yws01}. Furthermore, a role from reflection on the host galaxy scales has been recently discussed in \citet{yan21}. In RBS 1055, if a fraction of the Fe K$\alpha$ emission line is produced from material located light-years away from the X-ray source, this could echo the higher intrinsic flux  of the source observed in the past years.\\
\indent The total mid-infrared (MIR) luminosities of the source, obtained with the {\it Wide-Field Infrared Survey Explorer} (WISE), have been reported in \citet[][indicated as 2MASSi J1159410-195924]{ichikawa17}: $\log(\frac{{\rm L_{4.6\ \mu \rm m}}}{\rm erg\ s^{-1}})=45.75$, $\log(\frac{{\rm L_{12\ \mu \rm m}}}{\rm erg\ s^{-1}})=45.76$ and $\log(\frac{{\rm L_{22\ \mu \rm m}}}{\rm erg\ s^{-1}})=45.82$. The observed luminosity at $12\ \mu \rm m$ is in agreement with the ones estimated using the $L_{12\ \mu \rm m}$-L$_{2-10\rm\  keV}$ correlations from \citet{gandhi09} and \citet{asmus15}, within statistical uncertainties. In the latter work, local AGN ($z<0.3$) in the luminosity range $42<\log(\frac{{\rm L_{\rm 2-10\ keV}}}{\rm erg\ s^{-1}})<46$ were considered. The MIR colors can be also used to estimate the amount of obscuration along the line of sight, as discussed in \citep{prr22}. Using their Eq. (3), from the ratio ${\rm L_{22\ \mu \rm m}}$/$L_{4.6\ \mu \rm m}$=1.17 we obtain a column density value which is in agreement with the upper limit found in Sect. 3.2. However, the obscuring column is known to correlate with the ${\rm L_{\rm 2-10\ keV}/{\rm L_{12\ \mu \rm m}}}$ ratio \citep{ichikawa12,yan19}. Eq. (1) in \citep{prr22} allows us to estimate the 2-10 keV luminosity from the inferred upper limit N$_{\rm H}<5\times10^{20}$ cm$^{-2}$. We find $\log\Big(\frac{{\rm L_{\rm 2-10\ keV}}}{L_{12\ \mu \rm m}}\Big)\simeq-0.348$ i.e. ${\rm L_{\rm 2-10\ keV}\simeq2.6\times10^{45}}$ erg s$^{-1}$, which supports the scenario in which the intrinsic flux of the source is steadily decreasing throughout the years.\\
\indent The broad band emission of RBS 1055 can be well described in terms of the two coronae model. A warmer population of electrons (kT$_e=0.12^{+0.08}_{-0.03}$ keV), with a Thomson optical depth $\tau=30_{-10}^{+15}$, is responsible for the soft excess below $\sim1$ keV and for the optical/UV emission, while a second, hotter corona (kT$_e=30^{+40}_{-10}$ keV) with $\tau=3.0^{+1.0}_{-1.4}$ accounts for the high energy spectral shape. These values are in agreement with those discussed in \citet{pud18} for a wider sample of radio quiet AGN. Their sample includes only one source at a redshift $z$>0.4 (namely HB890405-123, at $z$=0.5725) and RBS 1055 therefore provides further confirmation of this model at high redshifts. Differently from \citet{mpb20} and \citet{upb20}, for the cases of Mrk 359 and HE 1143-1810, respectively, a contribution from relativistic reflection off the inner regions of the accretion disk is unlikely, since a broad Fe K$\alpha$ component is not found. We note that the inferred values for the high energy cutoff and for the temperature of the hot corona follow the E$_{\rm C}\simeq$(2-5)kT$_e$ trend \citep{mbm19}. \\
\indent As reported in Sec. \ref{sec:continuum}, by analizing the optical spectrum of RBS1055 we found a very high Balmer decrement for the BLR emission, which is indicative of reddening in this region.  An excess of optical reddening has been observed in several other AGN (\citealt{Barcons2003}, \citealt{Pappa2001}, \citealt{Carrera2004},\citealt{Corral2005}). The explanations suggested for this behaviour are related to (i) the presence of a dusty warm absorber and (ii) an intrinsic property of the BLR. The first hypothesis could partially explain our findings, as the presence of a warm absorber is marginally detected in RBS1055 (see Sec \ref{sec:xray}). The very high Balmer decrement could also suggest that case B recombination is not always valid in the BLR, and hence it could be an intrinsic property of the BLR, or it is the result of a peculiar geometry according to which our line of sight 
intercepts dust that reddens the BLR emission but not the optical continuum.

\begin{figure}
\centering
  \epsfig{file=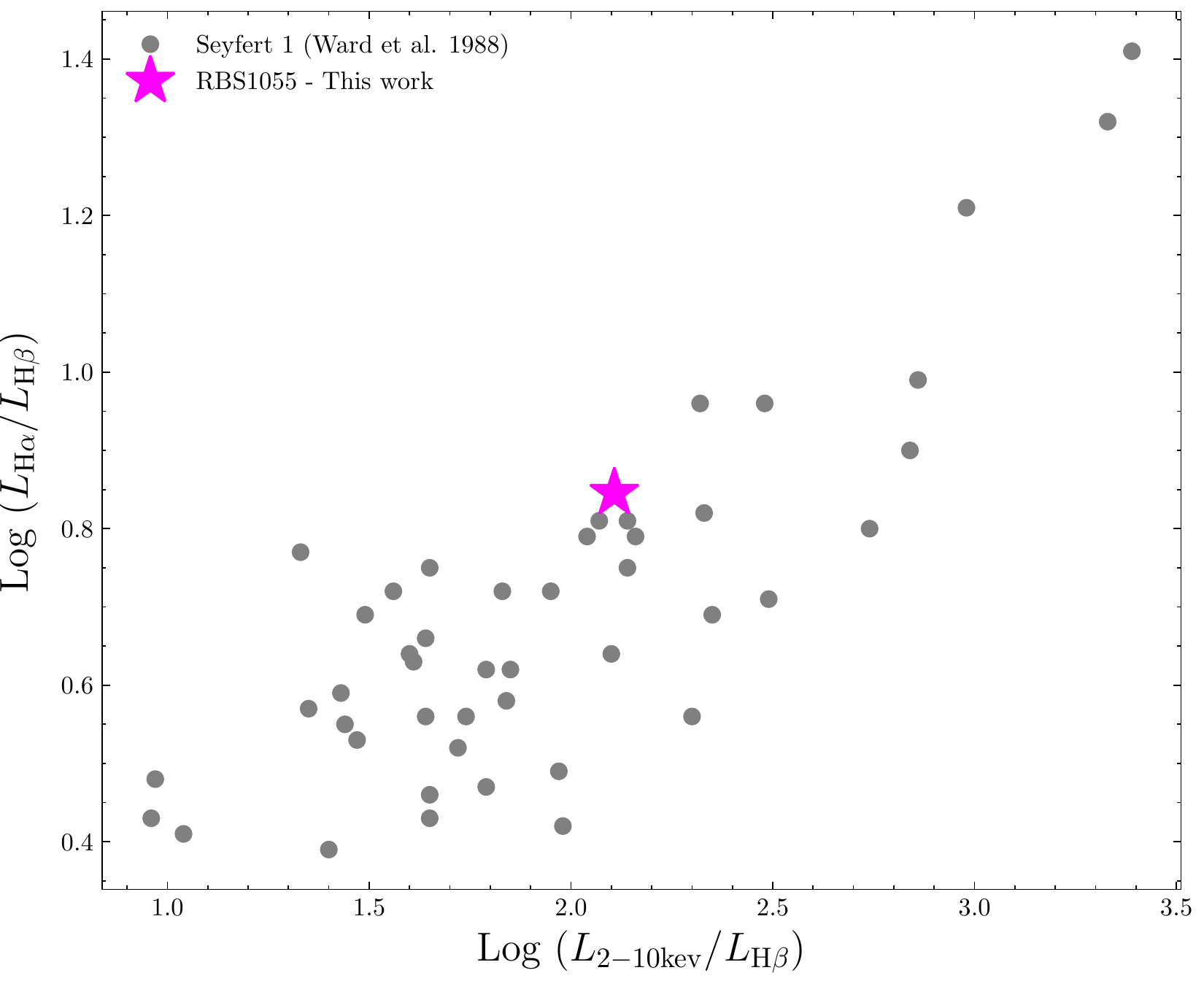, width=1.0\columnwidth}
  \caption{Correlation between Log ($L\rm_{H\alpha}$/$L\rm_{H\beta}$) and Log ($L\rm_X$/$L\rm_{H\beta}$). Magenta star represents RBS1055 values, while grey dots refer to Seyfert 1 sample from \cite{Ward1988}.}
  \label{fig:Lx_balmer}
\end{figure}

Such dust layers partially obscuring the BLR could be located between the BLR probed by the Balmer and the MgII lines, as the latter is only slightly redshifted ($\Delta \rm v_{MgII}$=480$\pm$100 km/s). To examine how our target relates to the bulk of the SDSS AGN population, we compared the measured H$\alpha$/MgII and H$\beta$/MgII line ratios of our target with those from the catalogue of \cite{Shen2011}, finding no difference with the majority of SDSS AGN, while the H$\alpha$/H$\beta$ ratio falls in the tail of the SDSS H$\alpha$/H$\beta$ ratio distribution. No peculiar properties are found when considering the relation between X-ray luminosity and Balmer lines, as shown in Fig. \ref{fig:Lx_balmer}. Here the Balmer decrement is shown as a function of the L$_X$/L$_{H\beta}$ ratio for a large sample of Seyfert 1 galaxies from \cite{Ward1988}, who found that the differences in the Balmer decrements are due to reddening rather than to an intrinsic property of the BLR. RBS1055 appears to be consistent with the trend reported in Fig. \ref{fig:Lx_balmer}, which supports the hypothesis of nuclear reddening. 

\section{Conclusions}
We analyzed and discussed the novel {\it NuSTAR} observation of RBS 1055 performed in March 2021 and the archival XMM-{\it Newton} pointings taken in July 2014. An optical spectrum of the source taken with the Double Spectrograph at the Palomar Observatory was also studied. Our main results can be summarized as follows:
\begin{itemize}
\item we confirm the presence of an intense Fe K$\alpha$ emission line at E=6.41$_{-0.01}^{+0.02}$ keV, with EW=55$\pm6$ eV. This measurement is $\sim3\sigma$ above the predicted value from the observed EW-L$_{\rm 2-10\ keV}$ relation \citep{bianchi07} and represents one of the few above L${}_{\rm 2-10\ keV}=10^{45}$ erg s$^{-1}$ with a robust Fe K$\alpha$ line. Once a toroidal model is considered to model the Compton reflection, a column density N$_{\rm H}=(3.2^{+0.9}_{-0.8})\times10^{23}$ cm$^{-2}$ is retrieved;\\
\item the primary nuclear continuum is well modelled with a cutoff power law with $\Gamma=1.70_{-0.05}^{+0.03}$, E$_{\rm C}>110$ keV and a  soft excess component is present in the pn/MOS spectra. We find that the two-coronae model \citep{ppm13,pud18}, well reproduces the broad band spectrum of RBS 1055, with temperatures kT$_e=0.12^{+0.08}_{-0.03}$ keV, kT$_e=30^{+40}_{-10}$ keV and Thomson optical depths $\tau$=30$_{-10}^{+15}$ and $\tau$=3.0$_{-1.4}^{+1.0}$ for the warm and hot coronal components, respectively;\\

\item the source also confirmed to have an extremely X-ray bright SED, with an inferred $\alpha_{\rm ox}=1.06$. This value is at the lower end of the observed $\alpha_{\rm ox}$ distributions \citep{mpz17, vat13};\\

\item  the optical spectrum reveals a likely peculiar configuration of our line of sight with respect to the nucleus, and the presence of a broad [O III] component, tracing outflows in the NLR, with a velocity shift $v=$1500$\pm100$ km s$^{-1}$, leading to a $\dot{M}_{\rm out}=25.4\pm1.5$ M$_{\odot}$ yr$^{-1}$ and $\dot{E}_{\rm kin}$/L$_{\rm Bol}$ $\sim$0.33\% (adopting the 5100\AA-based $L\rm_{Bol}$ value).

\end{itemize}

\section*{Acknowledgements}
 We thank the anonymous referee for their comments and suggestions, which greatly improved the paper. AM, GV, EP, SB, GL, and CV acknowledge support from PRIN MIUR project "Black Hole winds and the Baryon Life Cycle of Galaxies: the stone-guest at the galaxy evolution supper",  contract no. 2017PH3WAT. GV and EP
acknowledge financial support under ASI-INAF contract 2017-14-H.0.
We made use of {\sc Astropy},\footnote{\url{http://www.astropy.org}} a community-developed core {\sc Python} package for Astronomy \citep{astropy13, astropy18} and {\sc matplotlib} \citep{h07}. This research has made use of the {\it NuSTAR} Data Analysis Software (NuSTARDAS) jointly developed by the ASI Science Data Center (ASDC, Italy) and the California Institute of Technology (USA).

\bibliographystyle{aa}
\bibliography{sbs} 

\end{document}